RESEARCH ARTICLE

# An Interacting Wasserstein Gradient Flow Strategy to Robust Bayesian Inference

Felipe Igea[1]* and Alice Cicirello[1,2]

[1] Department of Engineering Science, University of Oxford, Parks Road, Oxford OX1 3PJ, UK

[2] Department of Engineering, University of Cambridge, Trumpington Street, Cambridge CB2 1PZ, UK

*Corresponding author: felipe.igea@eng.ox.ac.uk

## Abstract

Model Updating is frequently used in Structural Health Monitoring to determine structures' operating conditions and whether maintenance is required. Data collected by sensors are used to update the values of some initially unknown physics-based model's parameters. Bayesian Inference techniques for model updating require the assumption of a prior distribution. This choice of prior may affect posterior predictions and subsequent decisions on maintenance requirements, specially under the typical case in engineering applications of little informative data. Therefore, understanding how the choice of prior may affect the posterior prediction is of great interest.

In this paper, a Robust Bayesian Inference technique evaluates the optimal and worst-case prior in the vicinity of a chosen nominal prior, and their corresponding posteriors. This technique employs an interacting Wasserstein gradient flow formulation.

Two numerical case studies are used to showcase the proposed algorithm: a double-banana-posterior and a double beam structure. Optimal and worst-case prior are modelled by specifying an ambiguity set containing any distribution at a statistical distance to the nominal prior, less or equal to the radius. Examples show how particles flow from an initial assumed Gaussian distribution to the optimal worst-case prior distribution that lies inside the defined ambiguity set, and the resulting particles from the approximation to the posterior.

The resulting posteriors may be used to yield the lower and upper bounds on subsequent calculations used for decision-making. If the metric used for decision-making is not sensitive to the resulting posteriors, it may be assumed that decisions taken are robust to prior uncertainty.

## Impact Statement

Bayesian Inference may be significantly sensitive to assumptions about the priors chosen for the latent parameters of the structure, especially if due to some restrictions such as time constraints and cost, the number of observations available is limited. In these cases, the selection of priors may affect the resulting posteriors, and as a consequence the decisions about reliability, useful lifetime and maintenance of the structure are influenced. To address these limitations a Robust Bayesian Inference approach based on an interacting Wasserstein gradient flows has been developed in this paper. The method estimates the optimal and worst cases of priors, and calculates their corresponding approximations to the posterior that may be used as lower and upper bounds on subsequent calculations used for decision-making. If the resulting metric used for decision-making does not change significantly from using the lower to using the upper bound, it may be assumed that the decisions taken are robust to prior uncertainty.

# 1. Introduction

Numerical simulations are used in a multitude of scientific and engineering areas to study the behaviour of complex systems under various conditions. However, the resulting numerical 'observations' from those numerical simulations have to be cautiously used during the inference process when compared with the corresponding experimental observations, as otherwise unfaithful posterior estimates of the uncertain parameters will be produced (Hermans et al., 2021). It should also be mentioned that the observations employed in the inference should be informative, providing new information about the system under analysis. In this paper, a Bayesian inference approach is developed with the purpose of assessing the sensitivity of the posterior predictions with respect to (w.r.t.) uncertain priors.

When a model is chosen to generate data, the statistical model is just an approximation, and some errors are unavoidable (Dewaskar et al., 2023): data noise, assumption of model normality, incorrect assumptions about parameters, etc. For the inference of some model's parameters, significant impact may be produced by only small errors on the specification of the statistical model (Dewaskar et al., 2023). The traditional approach has been the use of a statistical model flexible enough to explicitly include the intricacies of the actual data (noises, outliers, etc.). As models characterising systems have grown in complexity, methodologies able to perform inference in these cases with intricate likelihood functions have been developed (Frazier, 2020). Unreliable approximations of the system's behaviour are mainly produced by not accounting for all plausible values of the observations that may be obtained from experiments. In engineering, this is of particular interest, as the number of experiments that may be run is limited due to the high cost incurred and time constraints. In these cases where the complexities of the likelihood are increased to improve the models' accuracy, and therefore, the reliability of the inferences, techniques such as: mixture models (Diakonikolas et al., 2020), nonparametric or semiparametric models (Lyddon et al., 2018), and models with heavier tails (Gonçalves et al., 2015) have been used as likelihood functions (Dewaskar et al., 2023). Nonetheless, the introduction of those methodologies to define the likelihood functions frequently lead to a set of new issues: higher numerical cost, definition of parameters, and harder interpretability. Although these techniques to define complex likelihood functions may improve the model's specification, some amount of inaccuracy is unavoidable (Dewaskar et al., 2023).

Current literature shows two main methodologies for robustifying inferences in the presence of model misspecification by focusing on the likelihood, and those can be broadly grouped into: (i) modified likelihood functions (Ghosh & Basu, 2016; Hooker & Vidyashankar, 2011), and (ii) distance-based estimation (Chérief-Abdellatif & Alquier, 2019; Matsubara et al., 2021). In the paper (Ghosh & Basu, 2016) an algorithm designed to produce robust Bayes estimators through modified likelihood functions, using the density power divergence is described. The method is designed to overcome the problems to manage outliers that may arise when the usual Bayesian estimator based on the ordinary posterior density is used. Another paper that uses the concept of modified likelihood functions is (Hooker & Vidyashankar, 2011). The method proposed is based on disparity theory, and it produces efficient and robust Bayesian inferences. Substituting the log likelihood by a suitably scaled disparity, the authors, using several examples, illustrate that robust inferences are obtained. The concept of distance-based estimation is used in the paper (Chérief-Abdellatif & Alquier, 2019) to produce a robust pseudo-posterior that is called the MMD-Bayes. MMD stands for Maximum Mean Discrepancy. The authors show that even for cases where noisy data and outliers are present, the MMD-Bayes posterior shows robustness to model misspecification. In their methodology, robustness is introduced through the substitution of the likelihood by the exponential of the MMD. In other words, this algorithm introduces in the Bayes equation the MMD-Bayes as a substitute loss function. Another example where robustness is sought through distance-based estimation is shown in (Matsubara et al., 2021). In (Matsubara et al., 2021), the loss function used in the Bayesian inference scheme is a Stein discrepancy. This method produces a robust posterior for cases where the likelihood is intractable/misspecified, and it generates a tractable posterior for Markov Chain Monte Carlo (MCMC).

However, when little data is available, Bayesian inference may be significantly sensitive to the assumptions of the prior for the latent parameters. In these cases where a limited number of observations is available, the choice of the prior may substantially affect the posterior obtained. In engineering, this may affect subsequent decisions such as those used to assess the reliability of a structure, its remaining useful lifetime, and whether a structure requires predictive maintenance. Therefore, a method able to quantify the robustness of the posterior prediction when the assumptions of the prior are changed, is of great interest. More specifically, if the optimal prior and the worst-case prior with respect to a metric could be obtained, the resulting posteriors may be used as lower and upper bounds on subsequent calculations used for decision-making. If the difference between the upper bound and lower bound found using the method is low for the metric used to support a decision, then it may be confirmed that the decision taken is robust to prior uncertainty. This is the focus of the present paper.

More specifically, it might not be possible to define exactly the prior for the latent variables. This type of situation could arise in the presence of conflicting opinions from experts. For those cases, we would like to explore how the approximation to the posterior might be affected by distributions that are in the neighbourhood of an assumed nominal prior, as this might have consequences on subsequent calculations. Therefore, it would be useful to develop a method that could determine the worst or most optimal distribution inside that neighbourhood of distributions in terms of a particular functional of interest.

In this paper, the problem of robustness to prior uncertainty in Bayesian Inference is dealt with by developing an interacting Wasserstein gradient flow combined with a so-called ambiguity set. An interacting Wasserstein gradient flow is derived to find: a) the best approximation to the posterior by minimising the Kullback-Leibler divergence (KL divergence) between the posterior and the approximation to the posterior, where the posterior is subject to change (due to the prior also changing). b) the (optimal or worst-case) prior that either minimises or maximises the KL divergence between the posterior and the approximation to the posterior. The proposed approach calculates the resulting optimal or worst-case prior by constraining the space of distributions to be explored using an ambiguity set. This ambiguity set is defined by a nominal distribution and all the distributions that lie within a specified value of a statistical distance, where both are assumed to be known. The robustness of the method is derived from this distance metric. A useful property of the Wasserstein distance is that distributions that do not share the same support may be investigated inside the ambiguity set (Kuhn et al., 2019). A particle-based interacting Wasserstein gradient flow Wasserstein-2 space algorithm is developed and the results from two numerical case studies are presented.

## 2. Robust Bayesian Inference Framework

The proposed Robust Bayesian Inference approach is based on the Wasserstein gradient flow formulation (Santambrogio, 2016). This method has been developed to deal with situations where the prior is uncertain, but it can be described by an ambiguity set (Bayraksan & Love, 2015). This is useful, as in some Bayesian Inference problems significant difficulties arise to define the priors of the latent parameters to be inferred. For example, when the suggestions of different experts about which priors should be used significantly differ. For the cases where the amount of observed data is limited, significant changes of posterior may be found for different choices of prior, and therefore, decisions to be taken for predictive maintenance may be affected. In these situations, an ambiguity set defined by a nominal prior, a statistical distance and a radius may be assumed, and the posterior resulting from identifying the optimal and the worst-case prior can be investigated by limiting the distribution space to priors within a statistical distance $\varepsilon$ of the nominal prior. In the next subsection, the concept of ambiguity set is defined.

### 2.1. Ambiguity Set

An ambiguity set is a set of distributions close to a reference distribution $p(\theta)$ with respect to some statistical distance $r$ (Bayraksan & Love, 2015). An ambiguity set is defined by the nominal distribution $p(\theta)$, a statistical distance $r$ and a radius $\varepsilon$. The ambiguity set is used to restrict the space of distributions that the prior could in theory take to solve the optimisation of the chosen functional. Figure 1, shows an ambiguity set that is centered at a nominal distribution $p(\theta)$ and contains any distribution $p^*$ within a statistical distance $r$ less or equal to $\varepsilon$, this may be expressed as:

$$A(\varepsilon, p) = \left\{ p^* : r\left(p^*(\theta) \| p(\theta)\right) \leq \varepsilon \right\} \tag{1}$$

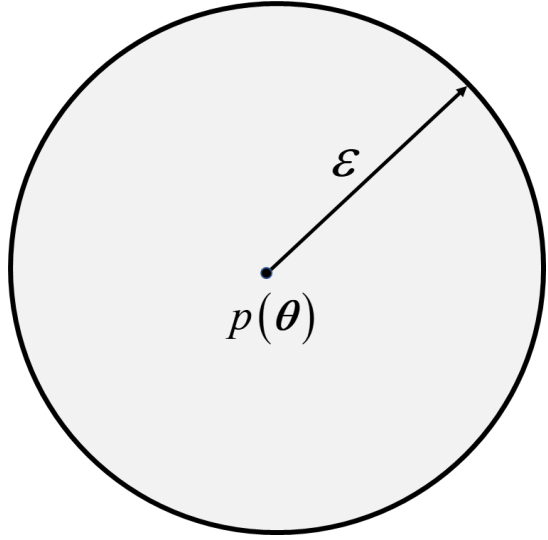

Figure 1. Ambiguity set centered at $p(\boldsymbol{\theta})$ with radius $\varepsilon$

When the ambiguity set is defined, two conditions must be met (Go & Isaac, 2022): $\int_\Omega p^*(\boldsymbol{\theta})d\boldsymbol{\theta} = 1$ and $r(p^*(\theta) \| p(\theta)) \leq \varepsilon$.

Any statistical distance may be used to define the ambiguity set, but care should be taken in choosing this distance, as which distributions lie inside that ambiguity set are defined by that statistical distance's properties. For example, if a phi divergence is used as the statistical distance in the ambiguity set, all distributions inside the ambiguity set must be absolutely continuous w.r.t. the nominal distribution (van Parys et al., 2017). However, if the 2-Wasserstein distance is used to define the ambiguity set, then the distributions that lie within the ambiguity set do not need to be absolutely continuous w.r.t. the nominal distribution (Kuhn et al., 2019). The use of the 2-Wasserstein distance also means that distributions that lie within the ambiguity set do not need to share the same support (Kuhn et al., 2019).

Depending on the information that the practitioner has available, the nominal distribution may be given by either an empirical distribution or a parametric distribution (e.g., Gaussian distribution) as shown in Figure 2. In Figure 2, $\hat{p}$ is a possible nominal distribution, $N$ is the number of data points, $\delta$ is the Kronecker delta function, $\xi$ is the parameter of the data, $\mathcal{N}(\hat{\mu}_N, \hat{\Sigma}_N)$ is a Gaussian distribution with sample mean $\hat{\mu}_N$ and sample covariance $\hat{\Sigma}_N$, obtained from $N$ data points.

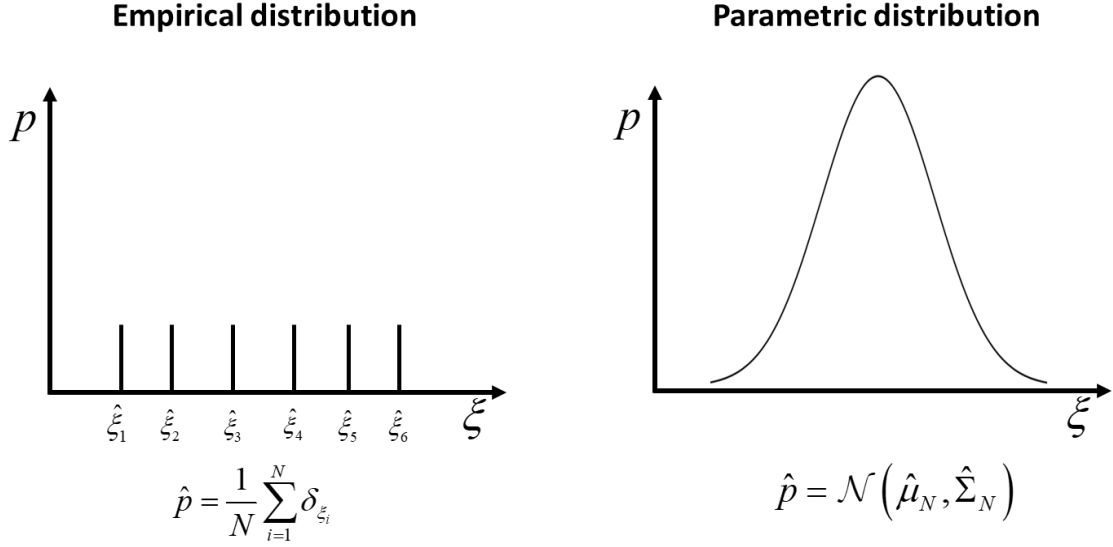

Figure 2. Nominal distributions: empirical vs. parametric distribution.

The chosen statistical distance for the ambiguity set is the 2-Wasserstein distance, as previously mentioned, this allows us to explore distributions that do not need to share the same support as the nominal distribution.

**2.2. Simultaneous optimization of approximated posterior and optimal or worst-case prior**

In this paper, we explore the problem of the simultaneous optimization of the approximation to the posterior and the optimal or worst-case prior by using an interacting Wasserstein Gradient Flow (WGF) scheme. The proposed approach differs from current Bayesian Inference WGFs based approaches, as it formulates a new problem that requires interacting WGFs for the simultaneous optimization of the chosen functional. This interacting WGF simultaneously obtains the best approximation to the posterior, and the optimal or worst-case prior that either minimises or maximises a certain functional. The ambiguity set is used to restrict the space of distributions that the prior could in theory take to solve the optimisation of the chosen functional $E(\rho(\theta), p(\theta))$. In this present paper, the min-max (or min-min) formulation problem that needs to be solved is:

$$\min_{\rho(\theta) \in \mathcal{P}(\Omega)} \min_{p^*(\theta) \in \mathcal{W}(p(\theta), p^*(\theta)) \leq \varepsilon} or \max E(\rho(\theta), p(\theta))$$

$$\text{where } E(\rho(\theta), p(\theta)) := \int \rho(\theta) \log\left(\frac{\rho(\theta)}{p(\theta) p(y_{obs} | \theta)}\right) \tag{2}$$

The distribution $\rho(\theta)$ is the approximation to the posterior, the likelihood distribution is $p(y_{obs}|\theta)$, the density $p(\theta)$ is the prior, and $\mathcal{W}$ is the 2-Wasserstein distance chosen to define the ambiguity set. The chosen functional $E(\rho(\theta), p(\theta))$ is the KL divergence between the unnormalized posterior $p(\theta, y_{obs})$ and the approximation to the posterior $\rho(\theta)$. This functional is chosen as it recently has been used to derive a WGF for Bayesian Inference (Y. Chen et al., 2023; Gao & Liu, 2020; Wang et al., 2022).

By using the properties of the logarithm, the functional in equation (2) can be rewritten as:

$$E(\rho(\theta), p(\theta)) := \int \rho(\theta) \log(\rho(\theta)) d\theta - \int \rho(\theta) \log(p(\theta)) d\theta - \int \rho(\theta) \log(p(y|\theta)) d\theta$$

The first term of the equation corresponds to the definition of entropy $\mathcal{H}$ w.r.t. the approximation to the posterior, therefore, the functional can be further expressed as:

$$E(\rho(\theta), p(\theta)) := -\mathcal{H}(\rho(\theta)) - \int \rho(\theta) \log(p(\theta)) d\theta - \int \rho(\theta) \log(p(y|\theta)) d\theta \quad (3)$$

The purpose of deriving an interacting Wasserstein gradient flow is to locate the pair of probability distributions $(\rho^*, p^*)$ that balances the simultaneous minimisation and maximisation (or minimisation) of the functional in equation (3). In other words, we are interested in finding simultaneously the distribution $\rho(\theta)$ that minimizes the KL divergence between the unnormalized posterior $p(\theta, y_{obs})$ and the approximation to the posterior $\rho(\theta)$, and the prior/s that minimizes/maximises the KL divergence between the unnormalized posterior $p(\theta, y_{obs})$ and the approximation to the posterior $\rho(\theta)$.

In numerous occasions, efforts have been made to prove the convergence of algorithms with interacting Wasserstein gradient flows to their global solution (Chizat & Bach, 2018; Mei et al., 2018), but these attempts generally require entropy regularization. The entropy regularisation is already included in the formulation of the functional shown in equation (3), where the first term regularises the partial differential equation of the WGF that minimises the KL divergence between the posterior and the approximation to the posterior. The second term in equation (3) serves as a regulariser of the WGF that minimises/maximises the KL divergence between the posterior and the approximation to the posterior to obtain the optimal or worst-case prior respectively. In this paper, it is assumed that the regularisers allow convergence to the pair of probability distributions that are sought. Proving convergence to this pair of probability distributions is still a problem currently under investigation, and not attempted to be solved in the current paper, the reader is referred to the Mixed Nash Equilibria literature for more details (Ding et al., 2023; Lin et al., 2019; Y. Lu, 2022).

### 2.3. Proposed algorithm and workflow

The proposed approach is schematically summarised in Figure 3, and it is composed of three main parts, the inputs, the simultaneous functional optimization and the outputs. The physics-based model (analytical, numerical or equivalent surrogate model) of the engineering system of interest, a nominal prior distribution on the unknown latent parameters with a specified radius, a statistical distance to define the ambiguity set, an assumed likelihood and measurements taken from the engineering system are needed as inputs. The main outputs, as shown in Figure 3, are the optimal or worst-case prior distribution and, consequently, the approximation to the posterior.

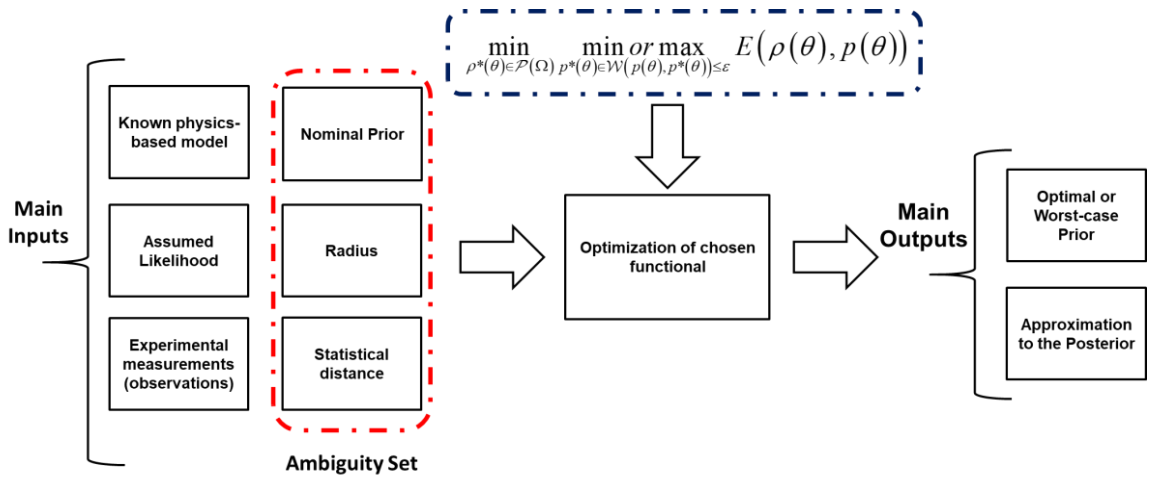

Figure 3. Main inputs, functional optimization and main outputs of the proposed approach.

In Figure 4, the elements of the optimization block shown in Figure 3 of the proposed approach are described. At first, we allow the approximation to the posterior to minimize the KL divergence between the posterior and the approximation to the posterior without changing the prior distribution. This is done by making the step size $\tau$ equal to zero of the WGF that results from either maximising or minimising the functional in equation (3) with respect to the prior for a prescribed amount of iterations $N_a$. The optimization to find the best approximation to the posterior is performed as follows. For the first iteration, $N_0$ initial particles are chosen at random (usually drawn from the nominal prior), and the same set of particles are used for both the initial prior $p_0(\boldsymbol{\theta})$ and the initial approximation to the posterior $\rho_0(\boldsymbol{\theta})$. At each iteration $i < N_a$, the physics-based model $PM(\boldsymbol{\theta})$ is run at the corresponding particle positions $\Theta_i^N$ of the approximation to the posterior. These numerical simulations at the particle positions $\Theta_i^N$ are then used to calculate the gradient of the logarithm of the likelihood $\nabla_{\boldsymbol{\theta}} \log p(\mathbf{y}_{\text{obs}} | \boldsymbol{\theta})$ at those respective locations. The gradient of the logarithm of the prior $\nabla_{\boldsymbol{\theta}} \log(p(\boldsymbol{\theta}))$, and the gradient of the logarithm to the approximation to the posterior $\nabla \log \rho_t(\boldsymbol{\theta})$, are approximated using a kernel density estimation approach, where the bandwidth is chosen using the median approach (Q. Liu & Wang, 2016). Using the equation (17), a new set of $N$ particles $\Theta_{i+1}^N \sim \rho_{i+1}(\boldsymbol{\theta})$ are obtained. This process is repeated until the iteration number reaches $i = N_a$, this is done to ensure that the approximation to the posterior has converged to the true posterior.

Once the prescribed number of iterations has been reached, the step size $\tau$ is allowed to be non-zero and positive, such that at every iteration $i \geq N_a$, a new set of prior particles $\boldsymbol{\theta}_{\text{prior},i+1}^N$ are obtained using the second equation (40). At this stage, both the approximation to the posterior and prior are updated using equation (40), such that the resulting new set of particles corresponds to independently and identically distributed samples from the distributions $\rho_{i+1}(\boldsymbol{\theta})$ and $p_{i+1}(\boldsymbol{\theta})$ of the next iteration number $i+1$.

Additionally, with the purpose of constraining the distribution to be optimised inside the ambiguity set, the 2-Wasserstein distance from the nominal prior to the prior at iteration $i$ is calculated at each iteration of the proposed method. If the distribution lies outside the ambiguity set, the distribution is discarded, and the size of the step in the particle flow algorithm is reduced until the distribution lies within the ambiguity set. In this way, the step size is controlled to restrict the prior within the radius of the Wasserstein ambiguity set. This is based on the assumption that the distribution that maximises or minimises the KL divergence between the actual posterior and the approximation to the posterior lies at the radius of the ambiguity set. Moreover, if a preset number $N_b$ of distributions are discarded when determining whether a distribution belongs in the ambiguity set, the distribution at iteration $i+1$ is reset to the distribution from an earlier iteration $i - N_c$ to avoid the optimisation getting trapped at one of the local optima. Convergence of the prior to the optimal or worst-case prior is assumed if the previously mentioned resetting occurs $N_{\text{reset}}$ times. In this case, the prior is no longer reset, and in a manner similar to the one defined at the beginning of the algorithm, an additional number of iterations $N_a$ are allowed, so the approximation to the posterior can converge. At this stage, the algorithm checks if the stopping criteria have been fulfilled, if it has not, a new iteration $i+1$ is started. The stopping criteria are set as: (i) the maximum number of allowed iterations $N_{\max}$ is reached; (ii) a maximum number of prior distributions $N_{\text{reset}}$ are reset to the distribution from an earlier iteration, and an additional number of iterations $N_a$ are allowed for the approximation to the posterior to converge.

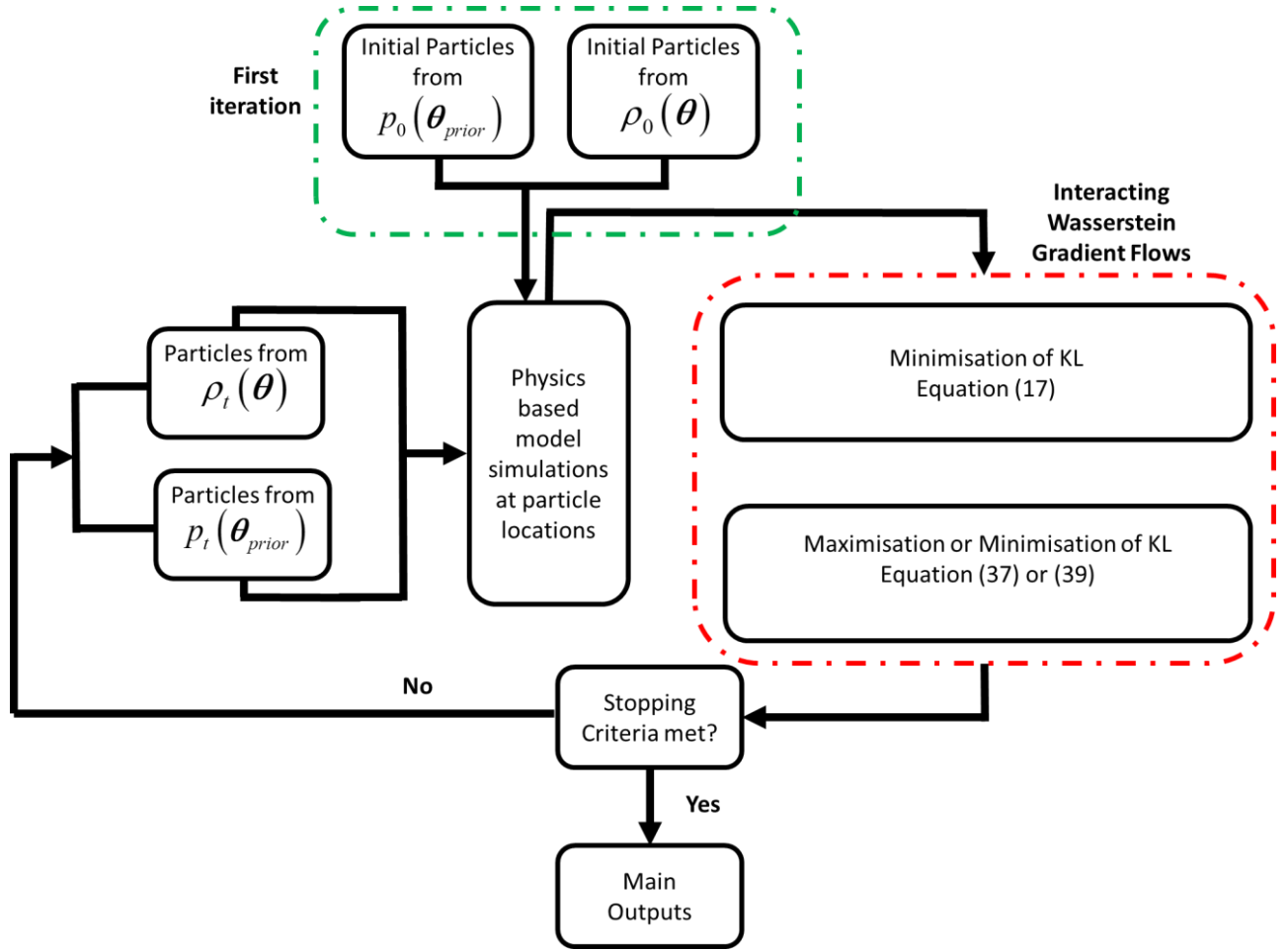

Figure 4. Pictorial description of simultaneous optimization of chosen functional.

A summary of the steps to be run for the proposed method are given below:

1. Calculate approximation to the posterior for the initial prior (by setting $\tau_t = 0$)
   a. Obtain $N_0$ initial particles from the prior, and approximation to the posterior.
   b. Calculate next set of particles of the approximation to the posterior using equation (17).
   c. Repeat from 1a) until iteration number reaches $i = N_a$.
2. Simultaneous optimization of equation (2) to calculate the approximation of the posterior and optimal or worst-case prior (allow $\tau_t > 0$):
   a. Calculate the next set of particles of the approximation to the posterior and prior using equations (17) and (37) or (39).
   b. Check if the prior lies outside the defined ambiguity set:
      i. if false, continue to 2c).
      ii. if true:
         1. reduce the time step $\tau_t$ until it is inside ambiguity set.
         2. check if the number of discarded distributions is less than $N_b$.
            a. if true, continue.
            b. if false, reset current prior particles to prior particles from iteration $i - N_c$.

3. check if the number of times prior distributions have been reset is less than $N_{reset}$.
   a. if true, continue.
   b. if false, skip to step 3.
  c. Repeat from 2a) until iterations reach $N_{max}$ and stop running the algorithm.
3. Calculate the approximation to the posterior for the final prior (by setting $\tau_t$ equal to zero, and allowing an additional number of iterations $N_a$):
   a. Calculate next set of particles of the approximation to the posterior using equation (17).
   b. Repeat from 3a) until iterations reach $N_{max}$ or the additional number of iterations is reached.

The system illustrated in Figure 5 is used to show the main results that would be obtained by using the proposed algorithm. A 1D mass-spring system with mass $m=1$ [kg], stiffness $k=1$ [N/m], and angular frequency $\omega = \sqrt{\frac{k}{m}}$ [rad/s] is studied. In this example, a Gaussian observational error $\sigma = \sqrt{0.1}$ is assumed when obtaining a numerical observation of the angular frequency $\omega_{obs} = \sqrt{\frac{k}{m}} + \zeta$, where $\zeta \sim \mathcal{N}(0,\sigma)$. It is also assumed, that the uncertain parameter is the spring stiffness $k = \theta$ [N/m]. The initial Gaussian prior (which is the same as the nominal prior of the ambiguity set) is assumed to be $p(\theta) = \mathcal{N}(1,0.1)$. Two different runs to obtain the optimal prior and the worst-case prior (and their corresponding approximation to the posterior) w.r.t. the chosen functional in equation (2) are shown in Figure 6. An ambiguity set with a radius $\varepsilon = 0.005$ is chosen. For the both optimal and worst-case prior, the step sizes in the interacting particle flow WGF algorithm cases are $\alpha = 3*10^{-3}$ and $\tau = 3*10^{-4}$. The values of $N_a$, $N_b$, $N_c$, $N_{reset}$ and $N_{max}$ used to run the algorithm are the same as the two numerical examples shown in Section 6 and they can be found in the introduction of that section. Also, in this example, the number of initial particles $N_0 = 100$ is chosen. The obtained probability density functions plotted in Figure 6 are calculated using the kde function in (*MATLAB*, 2020) with the standard options, and using 100 samples from their respective distributions.

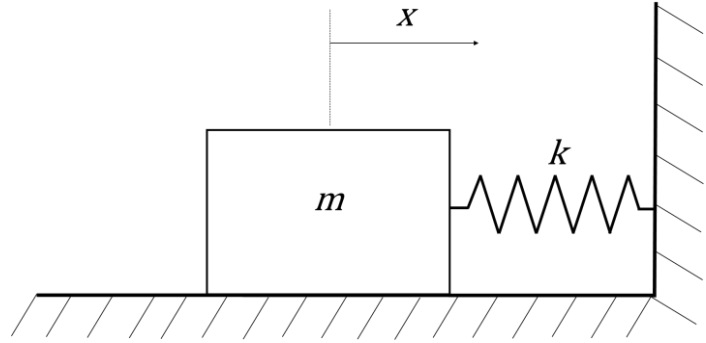

Figure 5. 1-Degree of freedom mass-spring system.

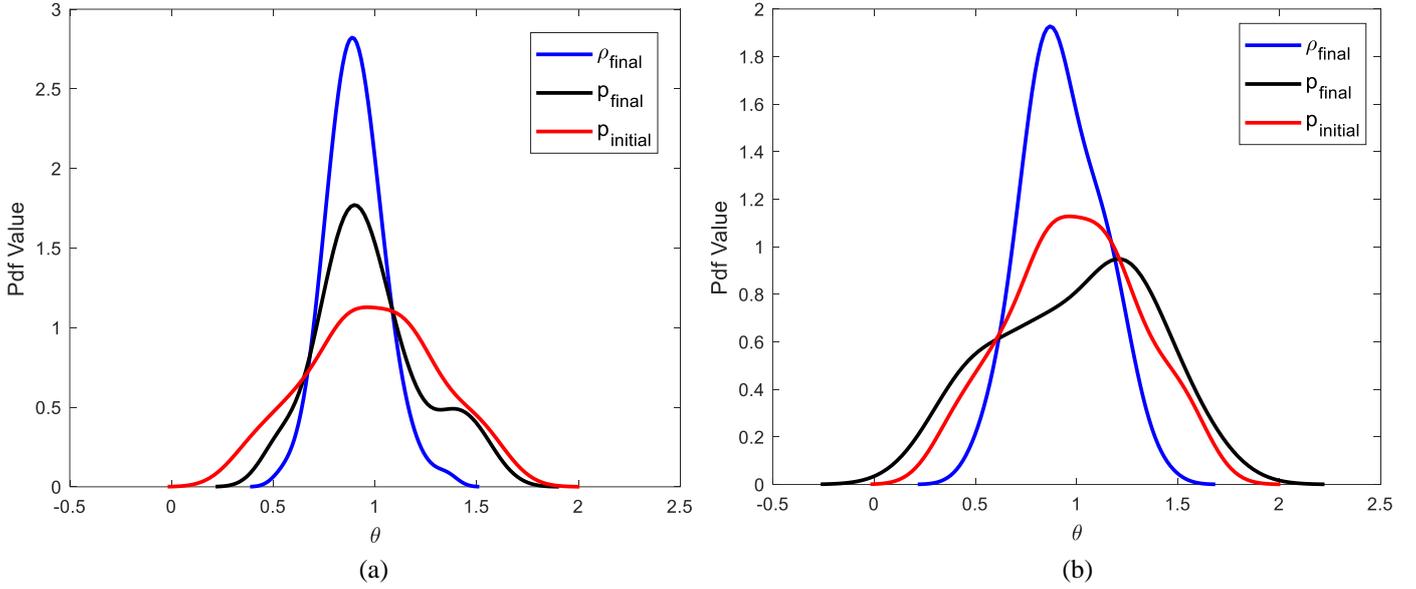

Figure 6. Kernel density estimates of the distributions for a 1D mass-spring system given an initial/nominal prior (red – initial prior; blue – final approximation to the posterior; black – final prior): (a) Optimal prior case (b) Worst-case prior case.

As expected, it can be seen on Figure 6, that the optimal prior case assigns higher probability density at regions of high posterior density, while the worst-case prior moves prior density away from regions of high posterior density. The optimal prior has its support reduced w.r.t. the initial prior, whilst for the worst-case prior its support is increased. A slight multimodality can be seen for the optimal prior, with its main mode at the same location as the only mode found for the approximation to the posterior. Both the optimal and worst-case prior distributions are non-Gaussian and non-symmetric, even though the initial prior was Gaussian, and therefore symmetric.

The following sections of this paper build upon the knowledge needed to understand the main concepts and algorithmic approximations required for the proposed approach.

### 3. Wasserstein Gradient Flow

In this paper, to be able to consider the optimisation of functionals with respect to probability measures, the Wasserstein gradient flow (WGF) (Santambrogio, 2015) concept is introduced. The WFG applies on a probability measures space where a 2-Wasserstein metric has been defined.

Let us first consider the functional $E(\rho)$, where $E : \mathcal{P}(\Omega) \to \mathbb{R}$ maps a probability measure to a real value, where the $\mathcal{P}(\Omega)$ is the space of probability measures on $\Omega \subset \mathbb{R}^D$, and $D$ is the number of dimensions.

To investigate the optimisation of the functional $E(\rho)$ as a Wasserstein gradient flow, the Jordan Kinderleher Otto (JKO) scheme (Ambrosio et al., 2005; Santambrogio, 2016) is used. The JKO scheme solves the variational problem by defining the time discretization of the diffusion process, for this discretization the approximate probability density, $\rho_{i+1}^{\tau}$ at the $i+1$ timestep is calculated:

$$\rho_{i+1}^{\tau} = \arg\min_{\rho} \left\{ E(\rho) + \frac{\mathcal{W}_2^2(\rho, \rho_i^{\tau})}{2\tau} \right\} \tag{4}$$

Where $\mathcal{W}_2$ is the 2-Wasserstein distance, $\tau > 0$ is the size of the timestep, and as the size of the timestep approaches zero $\tau \to 0$, the expression above converges to the exact Wasserstein gradient flow. The 2-Wasserstein distance (curve length between two distributions) is defined as (Santambrogio, 2015):

$$\mathcal{W}_2^2(\mu,\nu) = \inf_{\gamma \in \Gamma(\mu,\nu)} \int_{\Omega \times \Omega} \|\boldsymbol{\theta} - \boldsymbol{\theta}^*\|_2^2 \, \gamma(d\boldsymbol{\theta}, d\boldsymbol{\theta}^*) \tag{5}$$

Where, $\gamma$ is the deterministic coupling that minimises equation (5) and $\gamma$ is inside the set of all possible couplings or joint distributions $\Gamma(\mu,\nu)$ over $\boldsymbol{\theta}$ and $\boldsymbol{\theta}^*$, where $\mu$ and $\nu$ are the marginal distributions of $\boldsymbol{\theta}$ and $\boldsymbol{\theta}^*$. In the context of transport optimization, the calculation of the 2-Wasserstein distance can be interpreted as the transformation of elements in the domain $\mu$ to the domain $\nu$ at a minimum cost. Then, from this perspective, in the equation (5) of the 2-Wasserstein distance, $\|\boldsymbol{\theta} - \boldsymbol{\theta}^*\|_2^2$ is the transportation cost of $\boldsymbol{\theta}$ in $\mu$, to $\boldsymbol{\theta}^*$ in $\nu$ (Santambrogio, 2015). By defining the cost function $c$ as $\|\boldsymbol{\theta} - \boldsymbol{\theta}^*\|_2^2$, the equation (5) can be rewritten to:

$$\mathcal{W}_2^2(\mu,\nu) := \inf_{\mathcal{T}} \int_{\Omega} c(\boldsymbol{\theta}, \mathcal{T}(\boldsymbol{\theta})) d\mu(\boldsymbol{\theta}) \tag{6}$$

For the cases where there is a unique solution for the problem of minimum transportation cost from $\boldsymbol{\theta}$ in $\mu$, to $\boldsymbol{\theta}^*$ in $\nu$, the unique solution can also be expressed as a mapping $\mathcal{T}: \mathbb{R}^D \to \mathbb{R}^D$, that pushes elements $\boldsymbol{\theta}$ of the domain $\mu$ to the domain $\nu$ (C. Chen et al., 2018). The solution is unique when the marginal distribution of probability $\mu$ is absolutely continuous w.r.t. the Lebesgue measure (C. Chen et al., 2018).

If $\{\mu_t\}_{t \in [0,1]}$ is an absolutely continuous curve with finite second-order moments in the probabilistic space $\mathcal{P}(\Omega)$, then the changes of $\mu_t$ in that curve will be defined through investigation of $\mathcal{W}_2^2(\mu_t, \mu_{t+\tau})$. Studying the changes of $\mu_t$, is related to the original JKO problem (Ambrosio et al., 2005; Santambrogio, 2016) of the minimisation of the functional shown in equation (4). These changes can be described using a velocity field given by: $\boldsymbol{v}_t(\boldsymbol{\theta}) := \lim_{\tau \to 0} \frac{\mathcal{T}(\boldsymbol{\theta}_t) - \boldsymbol{\theta}_t}{\tau}$. This velocity field $\boldsymbol{v}_t(\boldsymbol{\theta})$ defines in $\mathcal{P}(\Omega)$ the gradient flow (Ambrosio et al., 2005):

$$\partial_t \mu_t + \nabla \cdot (\boldsymbol{v}_t \mu_t) = 0 \tag{7}$$

Solving the equation (7) requires finding a velocity field $\boldsymbol{v}(t)$ such that its flow agrees with $\lim_{\tau \to 0}(\boldsymbol{\theta}_\tau(t))$. The WGF can be shown to have a velocity field $\boldsymbol{v}(t)$ that minimises the functional $E(\rho)$, with the following form $\boldsymbol{v}(t) = -\nabla \frac{\partial E(\rho)}{\partial \rho}$ (Ambrosio et al., 2005), where $\frac{\partial E(\rho)}{\partial \rho}$ is called the first variation of $E(\rho)$ at $\rho$. Based on this, the WGF may be expressed as:

$$\partial_t \rho_t = -\nabla \cdot (\boldsymbol{v}_t \rho_t) = \nabla \cdot \left( \rho_t \nabla \frac{\partial E(\rho_t)}{\partial \rho_t} \right) \tag{8}$$

Therefore, to derive the WGF for the optimisation of the functional $E(\rho)$ the following requirements are introduced:

1. The first variation of the functional $E(\rho)$ with respect to the density $\dfrac{\partial E(\rho)}{\partial \rho}$ needs to be calculated.

2. A perturbation that follows the formal definition of a derivative in the probability space has to be introduced.

3. The probability $\rho$ is a probability density $\rho \in \mathcal{P}(\Omega)$ that has to be perturbed to $\rho + \varepsilon\chi$ which is also another probability density such that it also lies in the probability space $\mathcal{P}(\Omega)$, in this way $E(\rho + \varepsilon\chi)$ is well defined.

4. For all small $\varepsilon > 0$, both, the perturbed probability density is defined in the probability space $\rho + \varepsilon\chi \in \mathcal{P}(\Omega)$, and $\sigma = \rho + \chi \in \mathcal{P}(\Omega)$.

This can also be rewritten as $\rho + \varepsilon\chi = \rho + \varepsilon(\sigma - \rho) = \rho(1-\varepsilon) + \varepsilon\sigma$, where $\rho(1-\varepsilon) + \varepsilon\sigma \in \mathcal{P}(\Omega)$, as long as $\rho$ and $\sigma$ are also probabilities densities.

Now that the requirements have been introduced, the first variation of $E(\rho)$, $\dfrac{\partial E(\rho)}{\partial \rho}$ can be found, and it is given as (Ambrosio et al., 2005; Santambrogio, 2016):

$$\left.\frac{\partial}{\partial \varepsilon} E(\rho + \varepsilon\chi)\right|_{\varepsilon=0} = \int_{\Omega} \frac{\partial E(\rho)}{\partial \rho} \chi(\boldsymbol{\theta}) d\boldsymbol{\theta} \qquad (9)$$

for all $\chi = \sigma - \rho$. If a constant $C$ is added, $\int_{\Omega}\left(\dfrac{\partial E(\rho)}{\partial \rho} + C\right)\chi(\boldsymbol{\theta})d\boldsymbol{\theta}$, it can be found that the first variation may be defined uniquely only up to additive constants, as that second integral $\int_{\Omega} \chi(\boldsymbol{\theta})d\boldsymbol{\theta}$ includes the difference of 2 probability densities $\chi = \sigma - \rho$.

### 4. Wasserstein Gradient Flow for Bayesian inference

Approximations to the posterior can be obtained using many different methods. Recently, methods based on Variational Inference (VI) have been gaining popularity (Blei et al., 2017). These methods are based on the minimisation of the KL divergence between the posterior $p(\boldsymbol{y}_{obs}|\boldsymbol{\theta})$ and a probability density (usually parametric) defined inside a family of distributions $\mathcal{Q}$, to quantify the degree of dissimilarity between two distributions over the same domain:

$$\rho^* = \arg\min_{\rho \in \mathcal{Q}} KL(\rho \| p(\boldsymbol{\theta}|\boldsymbol{y}_{obs})) \qquad (10)$$

where the KL divergence is defined as:

$$KL(\rho \| p(\boldsymbol{\theta}|\boldsymbol{y}_{obs})) = \int_{\Omega} \rho \log\left(\frac{\rho}{p(\boldsymbol{\theta}|\boldsymbol{y}_{obs})}\right) d\boldsymbol{\theta} \qquad (11)$$

The approximation to the posterior is obtained by finding the member of the family and its respective hyperparameters that best minimise the KL divergence (Blei et al., 2017).

An alternative to VI would be to use the Wasserstein Gradient flow to define an iterative procedure that uses the set of data $\boldsymbol{y}_{obs}$ to update a chain of $\rho_n(\boldsymbol{\theta})$ with the purpose of approximating $p(\boldsymbol{\theta}|\boldsymbol{y}_{obs})$ given the minimisation of a suitable functional $E(\rho)$. In WGF, the optimisation of the functional can be solved by using equation (8). To be able to solve this equation, the velocity field $v(t)$ given the chosen functional is required. In a first analysis, it may be thought that as the posterior $p(\theta | y_{obs})$ is not known in advance, (because of the presence of the normalization constant $p(y_{obs})$), then the functional of equation (11) cannot be used to derive a WGF. But as the first variation of the functional is only uniquely defined up to additive constants, a simpler functional $E(\rho)$ where the posterior $p(\theta | y_{obs})$ is replaced for the unnormalized posterior $p(\theta, y_{obs})$ may be used (Gao & Liu, 2020). Therefore, the velocity field that results from replacing the posterior with the unnormalized posterior would be the same as the velocity field as in the functional in equation (11).

By obtaining a WGF of the functional $E(\rho)$, the partial differential equation can be solved to flow the approximation to the posterior $\rho_t(\theta)$ to its equilibrium $p(\theta | y_{obs})$ for the observed data. The dynamic system is defined by an initial density $\rho_0(\theta)$ that is given by the prior $p(\theta)$, and $\rho_\infty(\theta)$ tends to the posterior distribution, (Gao & Liu, 2020). In a more rigorous manner, in a manifold $M$ in the parameter space, a pushforward density $\rho_t(\theta) = T_t \# p(\theta) \in M$ is considered, where $\#$ is the push forward operator, and the best curve (under certain restrictions) $\rho_t$, that drives $\rho_0$ to $\rho_\infty$ has to be found (Gao & Liu, 2020).

The WGF of the chosen functional $E(\rho)$, may be performed by first calculating the first variation (where the bounds are omitted for clarity):

$$\frac{\partial}{\partial \varepsilon} E(\rho + \varepsilon \chi)\bigg|_{\varepsilon=0} = \frac{\partial}{\partial \varepsilon}\left[\int (\rho + \chi \varepsilon)\log((\rho + \chi \varepsilon))d\theta - \int (\rho + \chi \varepsilon)\log(p(\theta, y_{obs}))d\theta\right]_{\varepsilon=0} =$$

$$= \left[\int \chi \log(\rho + \chi \varepsilon)d\theta + \int (\rho + \chi \varepsilon)\frac{\chi}{(\rho + \chi \varepsilon)}d\theta - \int (\rho + \chi \varepsilon)\log(p(\theta, y_{obs}))d\theta\right]_{\varepsilon=0} =$$

$$= \int \chi \log(\rho)d\theta + \int \chi d\theta - \int \chi \log(p(\theta, y_{obs}))d\theta = \int \left(\log(\rho) + 1 - \log(p(\theta, y_{obs}))\right)\chi d\theta \quad (12)$$

The first variation of the functional $E(\rho)$ with respect to the density $\rho$ is then given by:

$$\frac{\partial E(\rho)}{\partial \rho} = \log(\rho) + 1 - \log(p(\theta, y_{obs})) \quad (13)$$

and the velocity field is:

$$v(t) = -\nabla \frac{\partial E(\rho)}{\partial \rho} = \nabla\left(\log(p(\theta, y_{obs})) - \log(\rho) - 1\right) = \nabla \log(p(\theta, y_{obs})) - \nabla \log(\rho)$$

(14)

If the first variation of the functional $E(\rho)$ is introduced into the continuity equation, the following equation is obtained (Wang et al., 2022):

$$\partial_t \rho_t = \nabla \cdot \left(\rho_t \left(\nabla \log(p(\theta, y_{obs})) - \nabla \log(\rho_t)\right)\right) \quad (15)$$

The KL WGF is an approximation in continuous time of the deterministic mean-field particle system called mean-field Wasserstein dynamics (Wang et al., 2022):

$$d\boldsymbol{\theta}_t = \left[\nabla \log\left(p\left(\boldsymbol{\theta}, \mathbf{y}_{obs}\right)\right) - \nabla \log\left(\rho_t\right)\right] dt \qquad (16)$$

The mean-field term is derived from the fact that the dynamics' evolution varies with the current density function $\rho_t$. The deterministic particle descent WGF may be obtained from the mean-field Wasserstein dynamics (Wang et al., 2022):

$$\boldsymbol{\theta}_{t+1} = \boldsymbol{\theta}_t + \alpha_t \left(\nabla \log\left(p\left(\boldsymbol{\theta}, \mathbf{y}_{obs}\right)\right) - \nabla \log\left(\rho_t\right)\right) \qquad (17)$$

The equation (17) represents one of the two particle discretisation WGFs equations needed for the simultaneous optimization of the chosen functional in equation (3) shown in Section 2. In equation (17), an approximation of $\nabla \log(\rho_t)$ is required, as no analytical expression is available. Many different methods may be used to obtain an approximation. In this paper, a kernel density estimate (KDE) approach is chosen and explained in subsection 5.1. It should be noted that the WGD follows a deterministic rule for the updating, and therefore the initial positions of the system determine the particle interactions and randomness.

## 5. Approximations in Wasserstein Gradient Flow for Robust Bayesian Inference

This section provides a more detailed explanation of some of the mathematical tools required for the application of the algorithm described in Figure 4.

### 5.1. Approximation to $\nabla_{\boldsymbol{\theta}} \log(\rho)$ from samples

When the velocity field $\mathbf{v}(t)$ is to be approximated, one of the difficulties that arises is the estimation of $\nabla \log \rho(\boldsymbol{\theta})$ (C. Liu et al., 2018). Only a finite set of samples $\{\boldsymbol{\theta}^{(i)}\}_{i=1}^N$ of $\rho(\boldsymbol{\theta})$ is known. However, a direct approximation of $\rho(\boldsymbol{\theta})$ using the empirical distribution $\hat{\rho}(\boldsymbol{\theta}) := \frac{1}{N}\sum_{i=1}^N \delta\left(\boldsymbol{\theta} - \boldsymbol{\theta}^{(i)}\right)$, where $\delta$ is the Dirac delta function is not possible. The reason why that direct approximation cannot be performed is because the WGF of the KL divergence at $\hat{\rho}(\boldsymbol{\theta})$ is not defined, consequence of $\hat{\rho}(\boldsymbol{\theta})$ not been absolutely continuous. Using the absolutely continuous approximated expression $\tilde{\rho}(\boldsymbol{\theta}) := (\hat{\rho} * K)(\boldsymbol{\theta}) = \frac{1}{N}\sum_{i=1}^N K\left(\boldsymbol{\theta}, \boldsymbol{\theta}^{(i)}\right)$ ("*" symbolizes convolution), the velocity field $\mathbf{v}(t)$ can be well-defined by smoothing $\hat{\rho}(\boldsymbol{\theta})$ through a smooth kernel $K$ on $\boldsymbol{\theta}$.

In this paper, the approximation of $\rho(\boldsymbol{\theta})$ is produced using the KDE $\tilde{\rho}(\boldsymbol{\theta})$, where $K\left(\boldsymbol{\theta}, \boldsymbol{\theta}^{(i)}\right) : \mathbb{R}^D \times \mathbb{R}^D \to \mathbb{R}$ is a given positive and differentiable kernel function, and the Gaussian kernel is used:

$$K\left(\boldsymbol{\theta}, \boldsymbol{\theta}^*\right) = \left(2\pi h\right)^{-\frac{N}{2}} \exp\left(-\frac{\|\boldsymbol{\theta} - \boldsymbol{\theta}^*\|_2^2}{2h}\right) \qquad (18)$$

$N$ is the number of samples used to define the kernel function $K\left(\boldsymbol{\theta}, \boldsymbol{\theta}^*\right)$, $h$ is the bandwidth and it is defined by $h = \text{med}^2 / \log(N)$, and $med$ represents the median of distances of the samples (Q. Liu & Wang, 2016).

When the KDE is used as an approximation of $\rho(\boldsymbol{\theta})$, the following expression may be used to calculate an approximation of $\nabla \log \rho(\boldsymbol{\theta})$ (Wang et al., 2022):

$$\nabla \log \tilde{\rho}(\boldsymbol{\theta}) = \frac{\nabla \tilde{\rho}(\boldsymbol{\theta})}{\tilde{\rho}(\boldsymbol{\theta})} = \frac{\sum_{i=1}^{N} \nabla_{\boldsymbol{\theta}} K(\boldsymbol{\theta}, \boldsymbol{\theta}^{(i)})}{\sum_{i=1}^{N} K(\boldsymbol{\theta}, \boldsymbol{\theta}^{(i)})} \quad (19)$$

The kernel chosen does not affect the solution of the Gradient Flow if the size of the ensemble tends to infinity (J. Lu et al., 2019). Nonetheless, the distribution of particles for a finite number of them, may not be unique. An alternative manner to explain this is that for different given kernels, that is, with different particle flows, different results (final positions in the state space of the particles) are obtained. However, for those kernels as their number of particles increases, the representation of the posterior probability density function (pdf) becomes more accurate.

## 5.2. Approximation to $\nabla_{\boldsymbol{\theta}} \log(p(\boldsymbol{\theta}, \boldsymbol{y}_{obs}))$

In this paper, two different ways to estimate the gradient of log likelihood are considered. The first one uses local estimations of that Jacobian matrix of the model's ensemble, whereas the second one uses gaussian processes. The first approach is only able to obtain estimates of the gradient of log likelihood at particle positions where the model has been run previously. However, the Gaussian Process approach is able to obtain estimates of the gradient of log likelihood at particle positions that have not been evaluated by leveraging on the prior assumptions and previous model runs. The choice of approach is usually based on the computational cost of dealing with the physics-based model involved.

In general $\nabla_{\boldsymbol{\theta}} \log(p(\boldsymbol{\theta}))$ can be calculated analytically, as most of the $\log(p(\boldsymbol{\theta}))$ chosen in Bayesian Inference are differentiable. However, if an analytical expression is not available, $\nabla_{\boldsymbol{\theta}} \log(p(\boldsymbol{\theta}))$ may be approximated using equation (19) as long as samples from the prior are available.

### 5.2.1. Gradient of log likelihood using ensemble method

Assuming a multivariate Gaussian likelihood, $p(\boldsymbol{y}_{obs} | \boldsymbol{\theta})$ can be written as:

$$p(\boldsymbol{y}_{obs} | \boldsymbol{\theta}) = \frac{1}{\sqrt{(2\pi)^d \det \Sigma}} \exp\left(-\frac{1}{2}(\boldsymbol{y}_{obs} - \boldsymbol{y}_{model})^T \Sigma^{-1} (\boldsymbol{y}_{obs} - \boldsymbol{y}_{model})\right) \quad (20)$$

In the above expression, $d$ refers to the dimensionality of the observation space (the number of observations), $\boldsymbol{y}_{model}$ and $\boldsymbol{y}_{obs}$ respectively are the $n \times 1$ vectors of simulated and observed states, and the inverse of the $n \times n$ error covariance matrix $\Sigma$ is denoted by $\Sigma^{-1}$.

By taking the gradient of the logarithm of the multivariate Gaussian likelihood, the below expression is obtained:

$$\nabla_{\boldsymbol{\theta}} \log p(\boldsymbol{y}_{obs} | \boldsymbol{\theta}) = \frac{1}{2} \nabla_{\boldsymbol{\theta}} \boldsymbol{y}_{model}^T \Sigma^{-1} (\boldsymbol{y}_{obs} - \boldsymbol{y}_{model}) \quad (21)$$

In equation (21), $\nabla_{\boldsymbol{\theta}} \boldsymbol{y}_{model}$ is a matrix of dimensions $n \times D$. The number of model parameters is denoted by $D$. The elements of the $\nabla_{\boldsymbol{\theta}} \boldsymbol{y}_{model}$ matrix are the partial derivatives of each simulated state (associated to rows 1, …, n) w.r.t. each parameter (associated to columns 1, …, D). The states are simulated introducing input parameters $\boldsymbol{\theta}$ into a computational model:

$$\boldsymbol{y}_{model} = PM(\boldsymbol{x}, \boldsymbol{\theta}) \quad (22)$$

The expression above assumes that the observed states are directly simulated by the model. If the Jacobian is defined as $J(\theta) = \nabla_\theta PM(x,\theta)^T$, the matrix of dimensions ($D \times n$), equation (21) can be rewritten as:

$$\nabla_\theta \log p(y_{\text{obs}} | \theta) = \frac{1}{2} J(\theta)^T \Sigma^{-1} (y_{\text{obs}} - y_{\text{model}}) \quad (23)$$

As a result, using the equation (23), the log likelihood gradient may be evaluated using local estimations of that Jacobian matrix. Computational difficulties arise during the evaluation of the Jacobian matrix $J(\theta)$ of dimensions $(n \times D)$, as the closed form of this matrix is frequently unavailable.

To solve the mentioned difficulty, an approach that consists of obtaining nonintrusive estimations of the Jacobian $J(\theta)$ may be taken (Ramgraber et al., 2021). The vector $\theta$, that has the parameters as elements, is perturbed in a small increment in each of its $D$ dimensions, and the Jacobian matrix $J(\theta)$ is approximated using the obtained two or three-points finite difference derivatives. This computational differentiation may produce very accurate results, but it becomes unpractical if the model has a high number of parameters $D$. If the ensemble size or number of particles is denoted $N$, and a set of local Jacobians is to be required, the model has to be run $(D+1)N$ times if two-points finite difference derivatives are used, or even more times $(2D+1)N$, if three-points finite difference derivatives are chosen (Ramgraber et al., 2021). Those numbers are well above the number of evaluations of the model that practitioners may consider affordable.

A technique that requires only $N$ model evaluations $PM(x,\theta)$, and it is able to produce the estimation of the Jacobian matrix $J(\theta)$, directly from the ensemble, may be found in (Ramgraber et al., 2021). This methodology makes use of the relative differences between particles:

$$\tilde{J}(\theta_r) = \frac{P}{N} \sum_{r=1}^{N} \frac{PM(\theta_r) - PM(\theta_s)}{\|PM(\theta_r) - PM(\theta_s)\|} \cdot \frac{\|PM(\theta_r) - PM(\theta_s)\|}{\|\theta_r - \theta_s\|} \cdot \frac{\theta_r^T - \theta_s^T}{\|\theta_r - \theta_s\|} \quad (24)$$

In the equation (24), $P$ is the rank expected for the Jacobian matrix $J(\theta)$, this expected rank is the smallest value between $N-1$ and $D$. Inside the summation symbol three fractions are found, in correlative order: the vector from particle $\theta_r$ to the particle $\theta_s$ (normalized), the scalar gradient between the observation and the parameter space, the normalized vector in parameter space. The equation (24) may be simplified as follows (Ramgraber et al., 2021):

$$\tilde{J}(\theta_r) = \frac{P}{N} \sum_{r=1}^{N} \frac{(PM(\theta_r) - PM(\theta_s))(\theta_r^T - \theta_s^T)}{\|\theta_r - \theta_s\|^2} \quad (25)$$

The factor $\frac{P}{N}$ external to the sum is made up of a correction factor $P$ to consider that the maximum possible contribution of each vector to the rank of the Jacobian is one, and a factor $\frac{1}{N}$ to account for an arithmetical average. For $N \to \infty$ and an isotropic arrangement of particles, the Jacobian in equation (25) should converge against the correct one (Ramgraber et al., 2021).

### 5.2.2. Gradient of log likelihood using Gaussian process

For cases where the physics-based model is expensive-to-evaluate, an approximation of the gradient of the log likelihood may be produced using a Gaussian process. This methodology allows the estimation of the gradient at particle positions where the physics-based model has not been evaluated.

Assuming that the likelihood function is given by a multivariate Gaussian with zero error mean and covariance $\Sigma$, the log likelihood function is:

$$\log p(\boldsymbol{y}_{obs} | \boldsymbol{\theta}) = -\frac{d}{2}\log(2\pi) - \frac{1}{2}\log(\det \Sigma) - \frac{1}{2}(\boldsymbol{y}_{obs} - \boldsymbol{y}_{model})^T \Sigma^{-1} (\boldsymbol{y}_{obs} - \boldsymbol{y}_{model}) \quad (26)$$

Focus is placed on the last term of the equation (26), as the gradient of the log likelihood function w.r.t. the parameter $\boldsymbol{\theta}$ only depends on that term. Consequently, the partially observed potential is modelled as (Dunbar et al., 2022):

$$V_L(\boldsymbol{\theta}) = \frac{1}{2}(\boldsymbol{y}_{obs} - \boldsymbol{y}_{model})^T \Sigma^{-1} (\boldsymbol{y}_{obs} - \boldsymbol{y}_{model}) \quad (27)$$

Where $V_L(\boldsymbol{\theta})$ is a Gaussian process $f \sim GP(0,k)$, and $k$ denotes a positive definite kernel on $\mathbb{R}^D$ that has been chosen according to the explanations below.

In this paper, $k$ is a Gaussian radial basis function kernel that has the form $k(\boldsymbol{\theta}, \boldsymbol{\theta}^*; \lambda, l) = \lambda \exp\left(-\frac{\|\boldsymbol{\theta} - \boldsymbol{\theta}^*\|^2}{2l^2}\right)$. In this expression, $l > 0$ denotes the kernel bandwidth, and $\lambda > 0$ is the amplitude of the kernel. A function $f$ is sought so that for some $\sigma > 0$, and for some noisy evaluations of the potential at the ensemble of points $\Theta_t = (\Theta_t^1, \ldots, \Theta_t^N) \in \mathbb{R}^{N \times D}$, then (Dunbar et al., 2022):

$$V_L(\Theta_t^i) = f(\Theta_t^i) + \sigma \xi^i, \quad \xi^i = (\xi^1, \ldots, \xi^N) \sim \mathcal{N}(0, I) \quad (28)$$

The mean function of the associated Gaussian process posterior for $f$ is (Rasmussen, 2003):

$$\mu(\boldsymbol{\theta}^*) = k(\boldsymbol{\theta}^*, \Theta) K(\Theta, \Theta)^{-1} V_L(\Theta) \quad (29)$$

and the expression for the variance function is (Rasmussen, 2003):

$$\sigma^2(\boldsymbol{\theta}^*) = k(\boldsymbol{\theta}^*, \boldsymbol{\theta}^*) - k(\boldsymbol{\theta}^*, \Theta) K(\Theta, \Theta)^{-1} k(\Theta, \boldsymbol{\theta}^*) \quad (30)$$

Where $K(\Theta, \Theta) = diag(\sigma^2) + k(\Theta, \Theta)$. Equations (31) and (32) express the well-defined gradient of the posterior mean (Rasmussen, 2003):

$$\mathbb{E}\left[\frac{\partial V_L(\boldsymbol{\theta}^*)}{\partial \boldsymbol{\theta}_d^*}\right] = \frac{\partial \mathbb{E}[V_L(\boldsymbol{\theta}^*)]}{\partial \boldsymbol{\theta}_d^*} = \frac{\partial k(\boldsymbol{\theta}^*, \Theta)}{\partial \boldsymbol{\theta}_d^*} K(\Theta, \Theta)^{-1} V_L(\Theta) \quad (31)$$

$$\nabla_{\boldsymbol{\theta}} V_L(\Theta) = \left.\frac{\partial k(\boldsymbol{\theta}^*, \Theta)}{\partial \boldsymbol{\theta}_d^*}\right|_{\boldsymbol{\theta}^* = \Theta} K(\Theta, \Theta)^{-1} V_L(\Theta) \quad (32)$$

Both the energy term $V_L(\theta)$ and the hyperparameters $(\sigma, \lambda, l)$ are updated at each iteration, and are calculated considering the new incoming data (Rasmussen, 2003).

### 5.3. Derivations of Wasserstein gradient flow equations for optimal or worst-case prior

In Section 4, the WGF for the case when the approximation to the posterior is made to vary to minimize the KL divergence between the posterior, and the approximation to the posterior has been derived. Now, the WGF that either maximises or minimises the KL divergence between the posterior and the approximation to the posterior with respect to the prior needs to be calculated. Currently, the interacting WGF has the following form:

$$\begin{cases} \partial_t \rho_t = \nabla \cdot \left( \rho_t \left( \nabla \log(p(\theta, y_{obs})) - \nabla \log(\rho_t) \right) \right) \\ \partial_t p_t(\theta) = \eta \left( -\nabla \cdot \left( p_t(\theta) \nabla \left( \frac{\partial E}{\partial p(\theta)} (\rho, p(\theta)) \right) \right) \right) \end{cases} \quad (33)$$

The first step is to calculate the first variation of the functional $E(\rho(\theta), p(\theta))$ with respect to the prior $p(\theta)$. When the optimal prior is of interest, this results in the minimization of equation (3), to obtain an expression of the first variation we first need to calculate the following:

$$\frac{\partial}{\partial \varepsilon} E(p(\theta) + \varepsilon \chi) \bigg|_{\varepsilon=0} = \frac{\partial}{\partial \varepsilon} \left[ \int \rho \log \rho \, d\theta - \int \rho \log(p(\theta) + \varepsilon \chi) d\theta - \int \rho \log(p(\theta | y_{obs})) d\theta \right]_{\varepsilon=0} =$$

$$= \left[ -\int \rho \frac{\chi}{(p(\theta) + \chi \varepsilon)} d\theta \right]_{\varepsilon=0} = -\int \frac{\rho}{p(\theta)} \chi d\theta \quad (34)$$

Now an expression of the first variation of the functional to be optimized can be obtained and it is given by:

$$\frac{\partial E}{\partial p(\theta)}(p(\theta)) = -\frac{\rho}{p(\theta)} \quad (35)$$

and the velocity field is:

$$v(t) = -\nabla \frac{\partial E}{\partial \rho}(\rho) = \nabla \left( \frac{\rho}{p(\theta)} \right) = \frac{p(\theta)}{p(\theta)} \nabla \left( \frac{\rho}{p(\theta)} \right) = \frac{\rho}{p(\theta)} \left( \nabla \log \rho - \nabla \log p(\theta) \right) \quad (36)$$

The resulting particle-based Wasserstein gradient flow, using an Euler discretisation, is given as:

$$\theta_{prior,t+1}^N = \theta_{prior,t}^N + \tau_t \left( \frac{\rho_t(\theta_{prior})}{p_t(\theta_{prior})} \left( \nabla \log p_t(\theta_{prior}) - \nabla \log \rho_t(\theta_{prior}) \right) \right) \quad (37)$$

If the maximization of the KL divergence is sought instead, this requires the calculation of the worst-case prior, and the resulting velocity field is given as the negative of the previously calculated velocity field:

$$v(t) = \frac{\rho}{p(\theta)} \left( \nabla \log p(\theta) - \nabla \log \rho \right) \quad (38)$$

Therefore, the resulting particle-based Wasserstein gradient flow using an Euler discretisation is given as:

$$\theta_{prior,t+1}^N = \theta_{prior,t}^N - \tau_t \left( \frac{\rho_t(\theta_{prior})}{p_t(\theta_{prior})} \left( \nabla \log p_t(\theta_{prior}) - \nabla \log \rho_t(\theta_{prior}) \right) \right) \quad (39)$$

Now that the particle based WGF for the minimisation or maximisation of the functional with respect to the prior has been derived, an interacting particle based WGF can be defined as follows:

$$\begin{cases} \theta_{t+1}^N = \theta_t^N + \alpha_t \left( \nabla \log \left( p_t(\theta, y_{obs}) \right) - \nabla \log(\rho_t) \right) \\ \theta_{prior,t+1}^N = \theta_{prior,t}^N \pm \tau_t \left( \frac{\rho_t(\theta_{prior})}{p_t(\theta_{prior})} \left( \nabla \log p_t(\theta_{prior}) - \nabla \log \rho_t(\theta_{prior}) \right) \right) \end{cases} \quad (40)$$

The resulting simultaneous equations (40) are composed of: a) the top equation which is the particle discretisation of the WGF that results from the minimisation of the KL divergence between the posterior and the approximation to the posterior w.r.t. the approximation to the posterior, and b) the bottom equation that results from either minimising or maximising the KL divergence between the posterior and the approximation to the posterior w.r.t. the prior. These simultaneous equations may be used to obtain the prior that either maximises or minimises the functional, and their resulting approximations to the posterior. For the case where the step size $\tau_t$ of the bottom equation in the simultaneous equation (40) is zero, the original particle based WGF for Bayesian Inference would be recovered, as this would mean the prior is static (it does not change with time).

**5.4. Density ratio estimation from samples**

The equations (37) and (39), require the calculation of the pdf of the $\rho_t(\theta_{prior})$ and the pdf of $p_t(\theta_{prior})$. This may be done for example using kernel density estimates. In this paper, rather than doing the direct estimation of the densities, the density ratio is calculated directly:

$$g(\theta_{prior}) = \frac{\rho_t(\theta_{prior})}{p_t(\theta_{prior})} \quad (41)$$

Numerous methods have been developed for the calculation of the density ratio in equation (41), the method chosen in this paper is the one called Relative unconstrained Least-Squares Importance Fitting (RuLSIF), and the interested reader can find it in (Yamada et al., 2011).

**6. Data and Numerical Models**

In this section, the proposed method is validated using two numerical examples. These two case studies have been selected to showcase the applicability of the proposed approach to deal with problems of different complexity, and an engineering case study is included. In the first example, the 2D double banana posterior problem (Detommaso et al., 2018) is used to show the resulting particles obtained from the optimal and worst-case prior, and also the resulting particles from the approximation to the posterior. In the second example, a double beam system is used to show the differences between the ensemble method and the Gaussian process to numerically estimate the gradient of the logarithm of the likelihood at the particle positions.

In both case studies, the number of initial samples is $N_0 = 100$ for the approximation to the posterior and also the prior, and those initial samples are picked from identically and independently distributed draws from the nominal prior. Each iteration of the algorithm uses the same number of particles ($N = 100$) and corresponds to evaluations of the physics-based model at the positions of those particles. As described in section 2, in the beginning of the method $\tau$ is set to zero until the number of iterations reaches $N_a = 50$.

The Gaussian kernel in equation (18), is used to produce the estimations of $\nabla \log \rho$ and $\nabla \log p(\theta)$, and the bandwidth is chosen using the median methodology.

If the distribution that is being optimised lies outside the ambiguity set, the size of the step in the particle flow algorithm is reduced to half until the distribution lies within the ambiguity set. Also, the distribution at iteration $i+1$ may be reset to a distribution from an earlier iteration $i - N_c$, where $N_c = 10$, if a preset number of distributions ($N_b = 5$) are discarded when determining whether a distribution belongs to the ambiguity set. The maximum number of prior distributions that are allowed to be reset is $N_{reset} = 2$, once this number is reached, an additional number of iterations $N_a$ are allowed. The total maximum allowed number of iterations $N_{max} = 400$.

### 6.1. Double banana posterior example

This first example is based on the paper (Detommaso et al., 2018) that results in a two-dimensional double-banana-shaped posterior distribution. The equation that defines the model used is given by the logarithmic Rosenbrock function used in (Detommaso et al., 2018):

$$PM(\boldsymbol{\theta}) = \log\left((1-\theta_1)^2 + 100(\theta_2 - \theta_1^2)^2\right) \tag{42}$$

The initial prior chosen is a standard multivariate Gaussian, $\mathcal{N}(0, I)$. The numerical observation used to update the prior knowledge is obtained by $y_{obs} = PM(\boldsymbol{\theta}_{true}) + \zeta$, where $\boldsymbol{\theta}_{true}$ is a random variable drawn from the assumed prior, the standard deviation of the observational error is σ = 0.3, and $\zeta \sim \mathcal{N}(0, \sigma^2 I)$.

For the ambiguity set, the nominal prior is chosen to be the same as the initial prior. The statistical distance used is the 2-Wasserstein distance, and a radius $\varepsilon = 0.05$ has been chosen.

Using the algorithm inputs described above, the interacting Wasserstein gradient flows are used to find the resulting distributions for two different cases: a) the optimal prior and its resulting approximation to the posterior and b) the worst-case prior and its resulting approximation to the posterior. In this example, the ensemble method described in section 5.2.1 is used to calculate an approximation to the gradient of the log likelihood at the particle positions to be evaluated.

The step sizes in the interacting particle flow WGF algorithm for the optimal prior case are $\alpha = 3*10^{-3}$ and $\tau = 1.5*10^{-3}$. For the worst-case prior case, the step sizes in the interacting particle flow WGF algorithm are $\alpha = 3*10^{-3}$ and $\tau = 3*10^{-4}$.

In this numerical case two different subcases are run, Figure 7 to Figure 13 correspond to the situations when the optimal prior and its approximation to the posterior are calculated. Figure 14 to 16 correspond to the situations when the worst-case prior and its approximation to the posterior are calculated.

In Figures 7 and 8 respectively is shown, for each iteration, the positions of the particles from the optimal prior and from the approximation to the posterior. In both plots of Figure 7, it may be seen that after iteration $i = N_a$ the inner particles of the prior initially tend inwards, i.e., to the direction of smaller absolute values of parameters $\theta_1$ and $\theta_2$, this is due to the fact that the prior tries to move the closest particles to the positions of the particles of the approximation to the posterior. It can also be seen that the prior particles after a number of approximately 150 iterations do not change much of position, this occurs because of the step size decrease performed with the purpose of constraining the prior inside the defined ambiguity set. Figure 8 illustrates how the particle positions of the approximation to the posterior start moving into the regions of higher probability density. After iteration $i = N_a$, the particles' positions of the approximation to the posterior concentrate even more into regions of high probability density due to the prior having a greater effect on the positions of the particles.

Figures 9 and 10 show respectively, for each iteration, the values of the gradient of the logarithm of the prior and the gradient of the logarithm of the likelihood at the particles $\Theta_i^N$ positions. Figure 9 shows how the values of the gradient of the logarithm of the prior at the particle positions of the approximation to the posterior start to decrease as the prior particles start concentrating around the particles of the approximation to the posterior. As the iterations progress, the values of the gradient of the logarithm of the prior at the approximation of the posterior particle positions start decreasing, this is because the particles of the prior become closer to the particles of the approximation to the posterior. This also means that the particles of the approximation to the posterior are becoming closer to regions of high prior density as the iterations progress. Figure 10 shows that during the initial iterations, high absolute values of the gradient of the logarithm of the likelihood may be found. This happens because during the initial stages of the algorithm there are particles that are still distant from the regions of high likelihood density. After around twenty to thirty iterations the values concentrate in a more defined region, even though some occasional extreme values can still be found.

In Figure 11, the initial particle positions (where the prior and approximation to the posterior particles are the same, shown in red), the final particle positions of the prior (black) and the approximation to the posterior (blue) can be seen. As expected, the final positions of the particles from the approximation to the posterior are shown to resemble the double-banana posterior in (Detommaso et al., 2018). It can also be observed that most of the final positions of the particles from the prior (optimal prior) are near the particles of the approximation to the posterior, and a smaller number of particles lie close to the initial prior particles. This means that the optimal prior assigns a high probability to the region close to the approximation to the posterior, and a lower probability to the outer particles far from the approximation to the posterior density.

Figure 12 shows a quiver plot, also known as vector plot, that is produced by the generic function `quiver` in (*MATLAB*, 2020). The scaling of the quiver function's default setting is chosen to prevent arrow length overlap. In this plot, the gradients of the logarithm of the prior, and the logarithm of the likelihood at the particle's positions from the approximation to the posterior in the final iteration are plotted. The gradients of the logarithm of the prior at the final prior particle positions are also shown.

Figure 13 illustrates the 2-Wasserstein distances at each iteration. Three plots can be found. The following distances at each iteration are plotted: the first is from the initial prior to the approximation of the posterior; the second from the approximation to the posterior and the prior; and the third from the initial prior to the prior.

In Figures 14 and 15 respectively, the positions of the particles from the worst-case prior, and from the approximation to the posterior, are shown for each iteration. In Figure 14 (a and b), it may be seen that the inner particles of the worst-case prior tend to move outwards to the direction of higher absolute values of parameters $\theta_1$ and $\theta_2$ as more iterations occur. Figure 14 also illustrates that after approximately 200 iterations the prior particles do not change much their positions. This is due to the decreasing size of the time step that is introduced with the purpose of limiting the prior inside the ambiguity set. In a similar manner to what occurs for the optimal prior case, in Figure 15 it can be observed that the particle positions of the approximation to the posterior also move to areas of higher probability density as iterations advance.

The final particle positions of the worst-case prior (black), initial particle positions (where the prior and approximation to the posterior particles are the same, shown in red), and its approximation to the posterior (blue) can be seen in Figure 16. As anticipated, the layout of the final positions of the particles from the approximation to the posterior takes a shape similar to the one shown by the double-banana posterior in (Detommaso et al., 2018). The worst-case prior assigns a higher density to areas of a low posterior density and vice versa. In a manner consistent with the previous statement, Figure 16 also shows that most of the final positions of the particles from the prior (worst-case prior), are positioned away from the final positions of the approximation to the posterior.

A direct comparison of the optimal prior and worst-case prior in the form of scatter plots and histograms of the latent variables is found on Figure 17. Figure 17 has been produced using the `plotmatrix` function from (*MATLAB*, 2020). It can be clearly seen that the optimal and worst-case prior differ from the initial prior and are no longer Gaussian. A very similar support can be seen of the optimal prior w.r.t. worst-case prior.

Scatter plots and histograms can also be found on Figure 18. For the cases where the optimal prior and worst-case prior have been estimated, the scatter plots and the histograms of the latent variables of the resulting approximation to the posterior are plotted. Very small differences are found when comparing between the resulting approximations to the posteriors. This is a consequence of the small sensitivity of the posterior to changes of the considered uncertain prior.

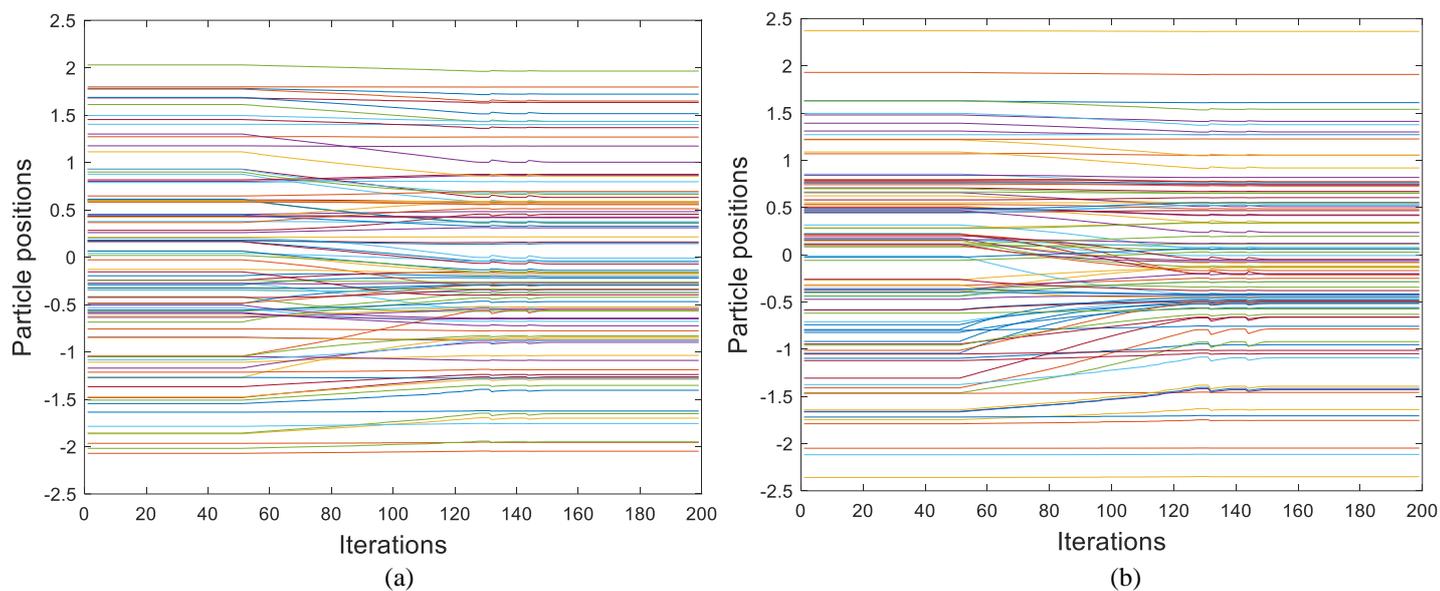

Figure 7. Optimal prior particle positions at different iterations: a) particle positions at $\theta_1$; b) particle positions at $\theta_2$.

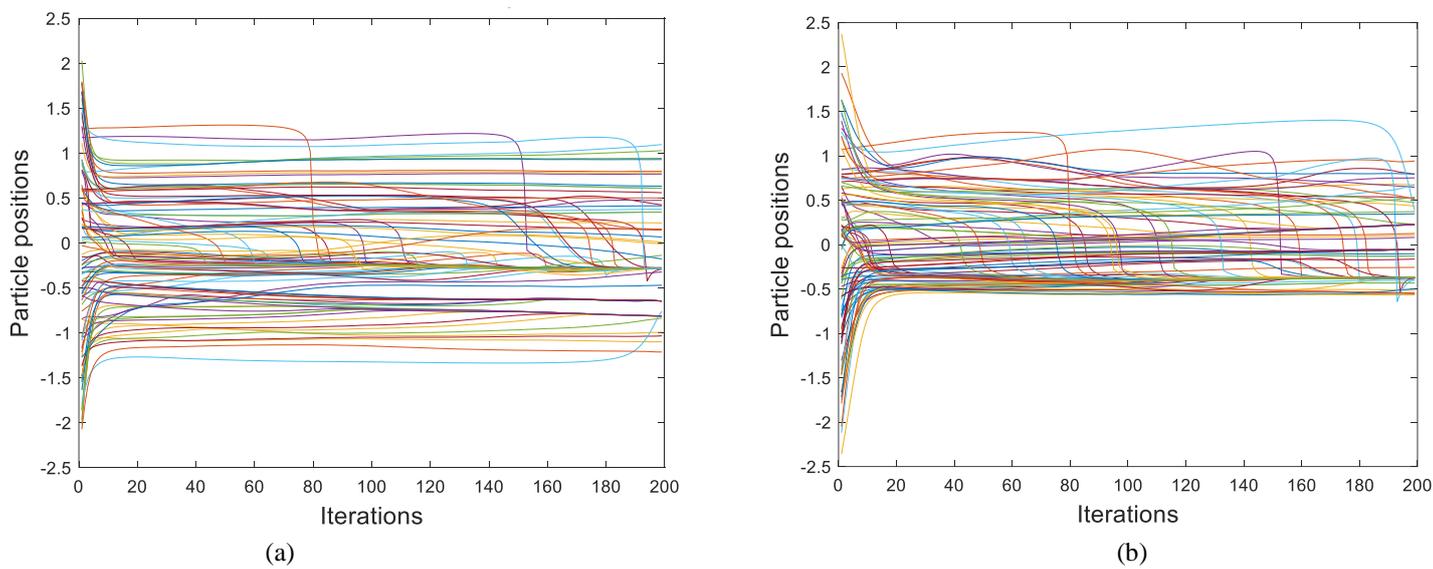

Figure 8. Approximation to posterior particle positions at different iterations: a) particle positions at $\theta_1$; b) particle positions at $\theta_2$.

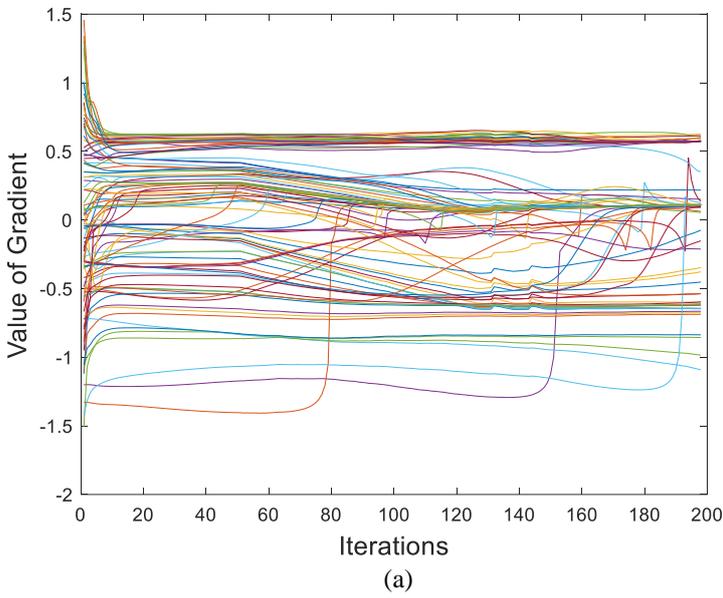 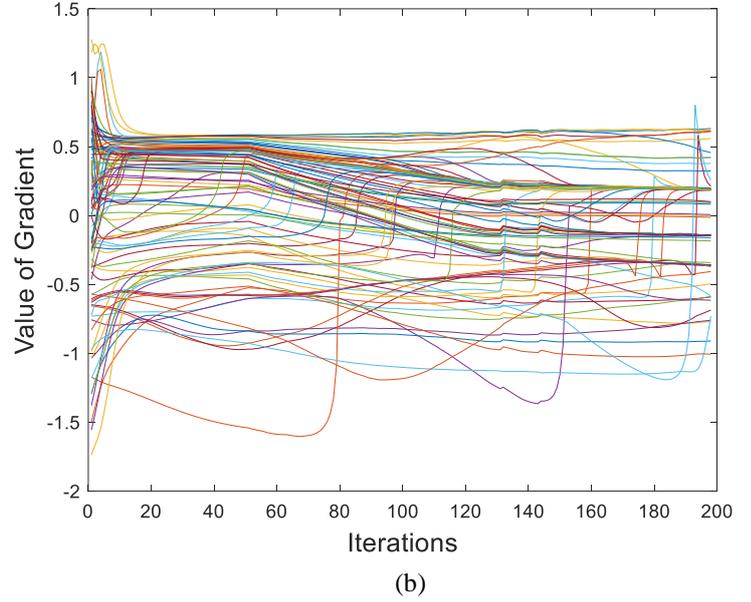

Figure 9. Gradient of log prior at different iterations and at particle positions $\Theta_i^N$ w.r.t.: a) latent parameter $\theta_1$; b) latent parameter $\theta_2$.

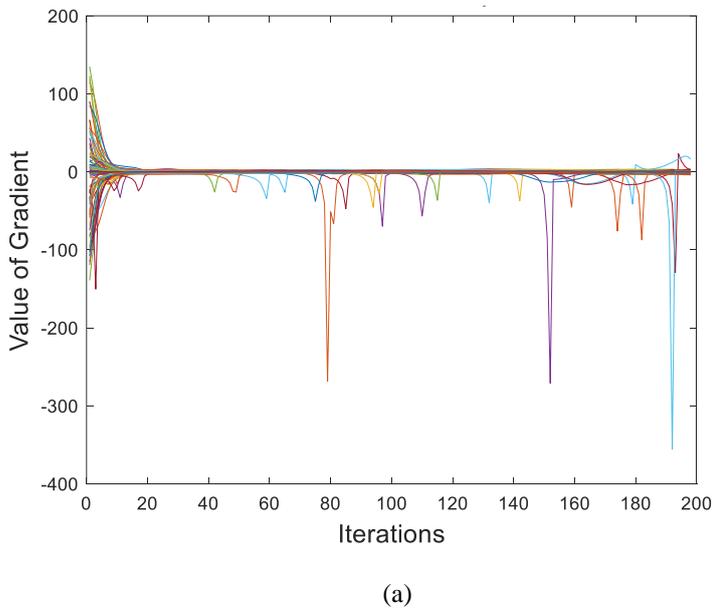 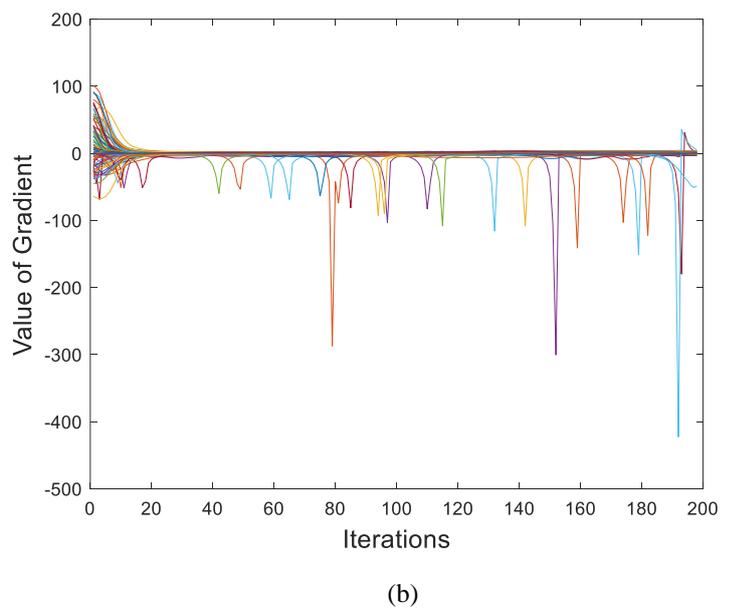

Figure 10. Gradient of log likelihood at several iterations and at particle positions $\Theta_i^N$ w.r.t.: a) latent parameter $\theta_1$; b) latent parameter $\theta_2$.

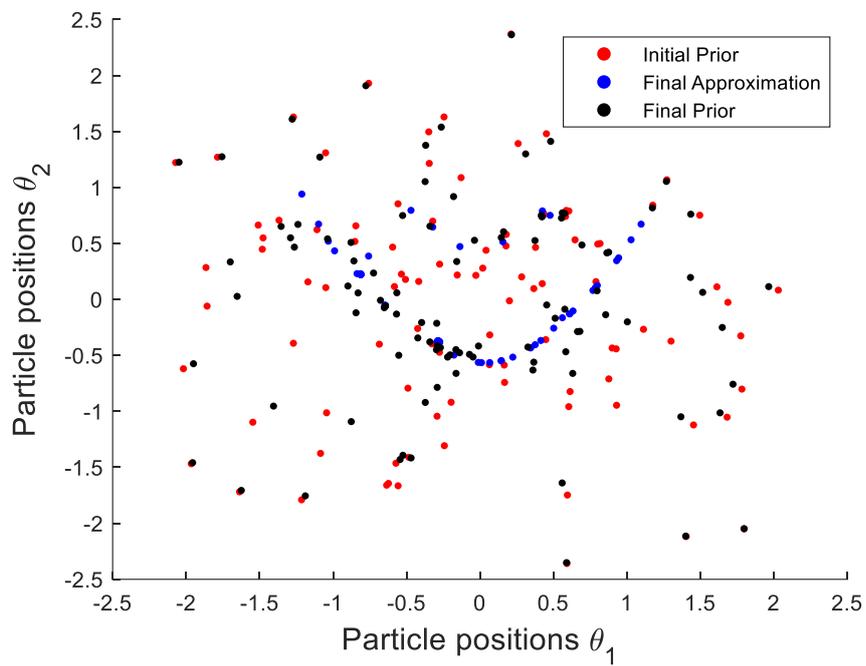

Figure 11. Initial prior, final approximation to the posterior and final prior particle positions.

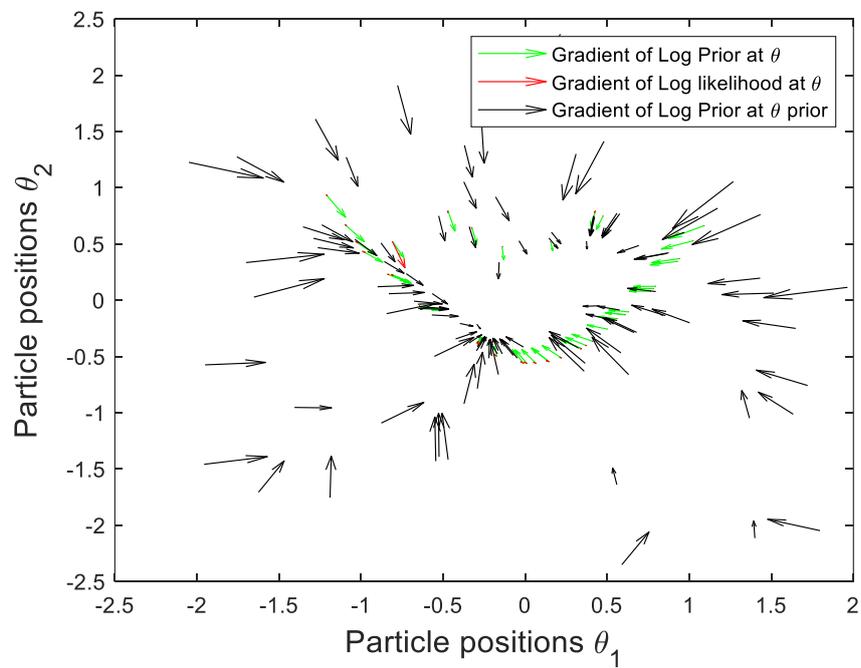

Figure 12. Gradient/Quiver Plot of log prior and log likelihood.

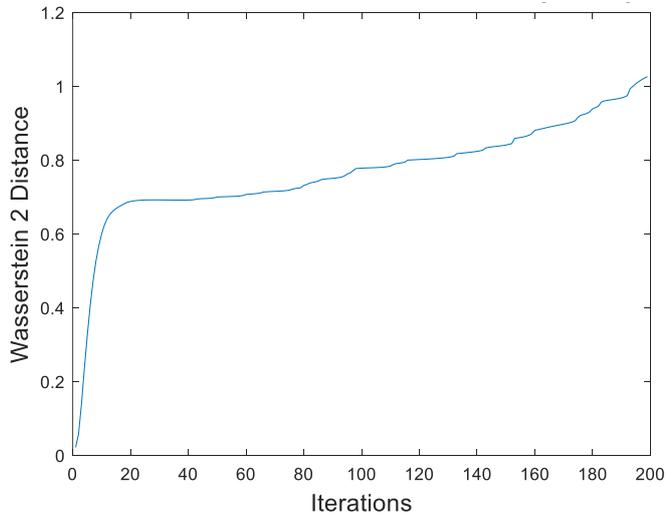
(a)

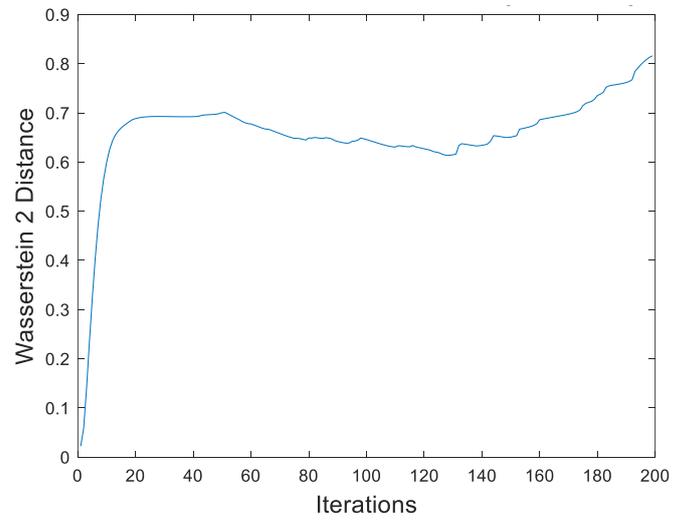
(b)

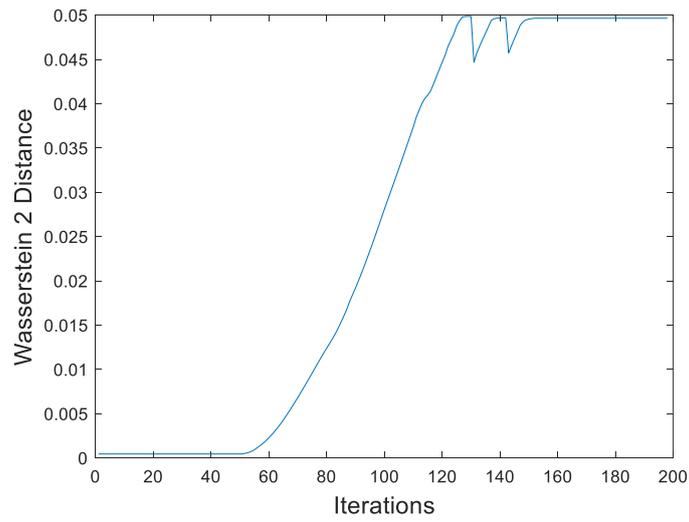
(c)

Figure 13. 2-Wasserstein distance at different iterations $i$ between: a) initial prior and approximation to posterior; b) approximation to posterior and prior; c) initial prior and prior.

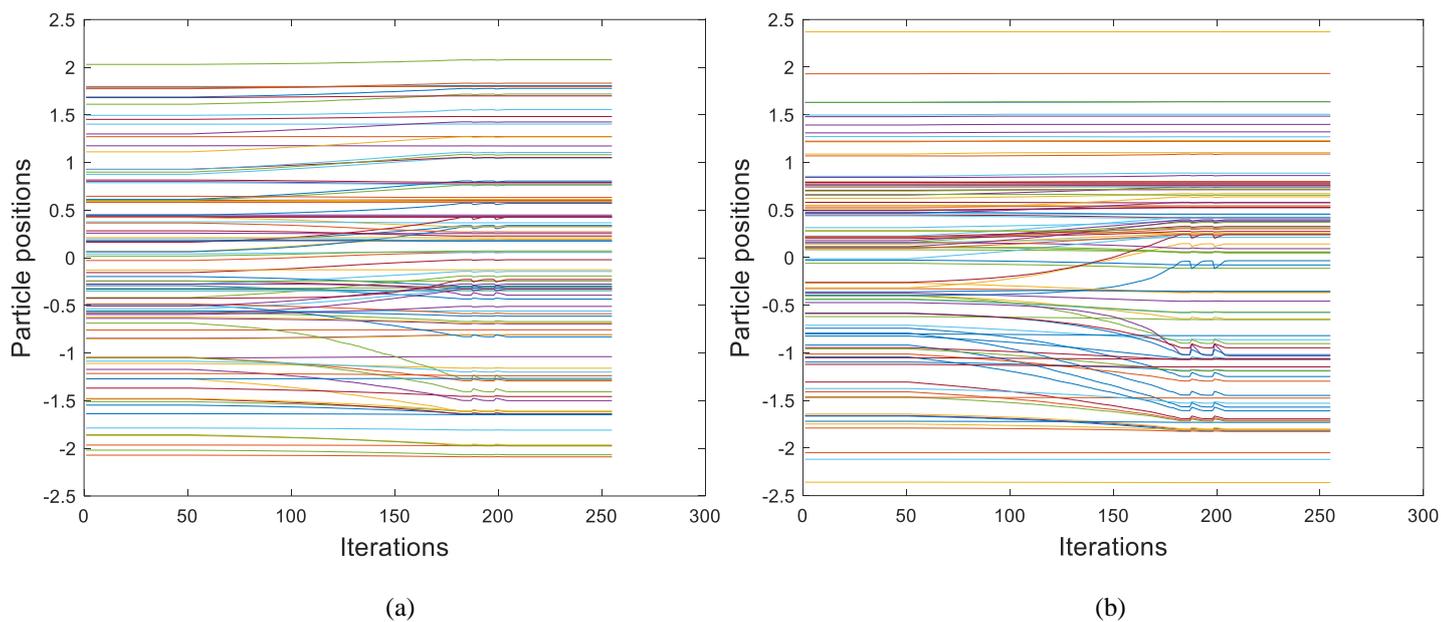

Figure 14. Worst-case prior particle positions at different iterations: a) particle positions at $\theta_1$; b) particle positions at $\theta_2$.

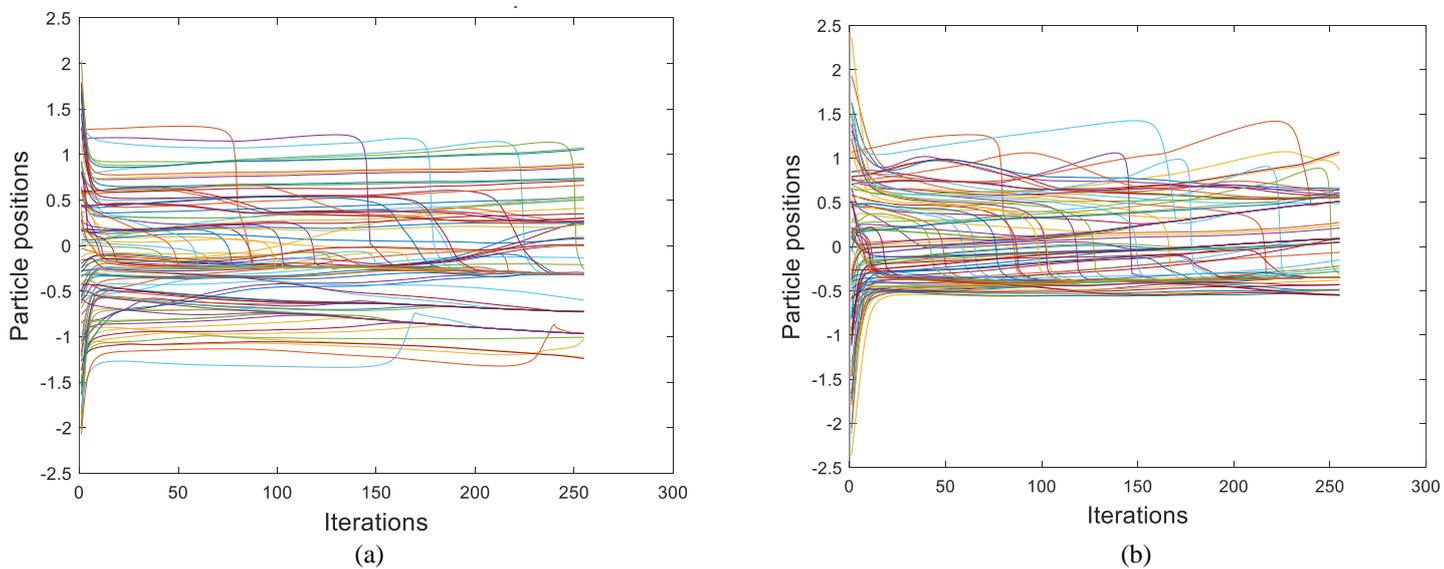

Figure 15. Approximation to posterior particle positions at different iterations: a) particle positions at $\theta_1$; b) particle positions at $\theta_2$.

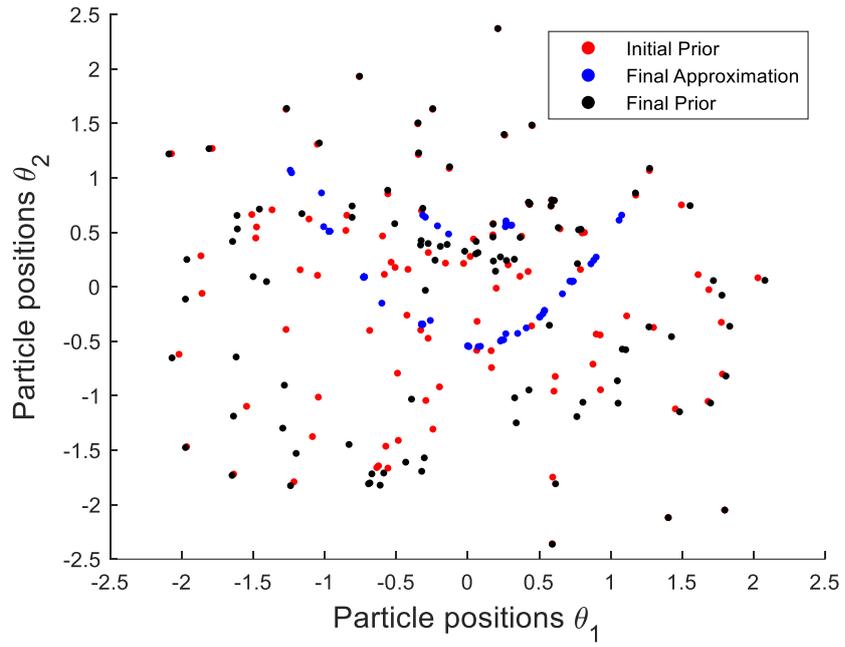

Figure 16. Initial prior, final approximation to the posterior and final worst-case prior particle positions.

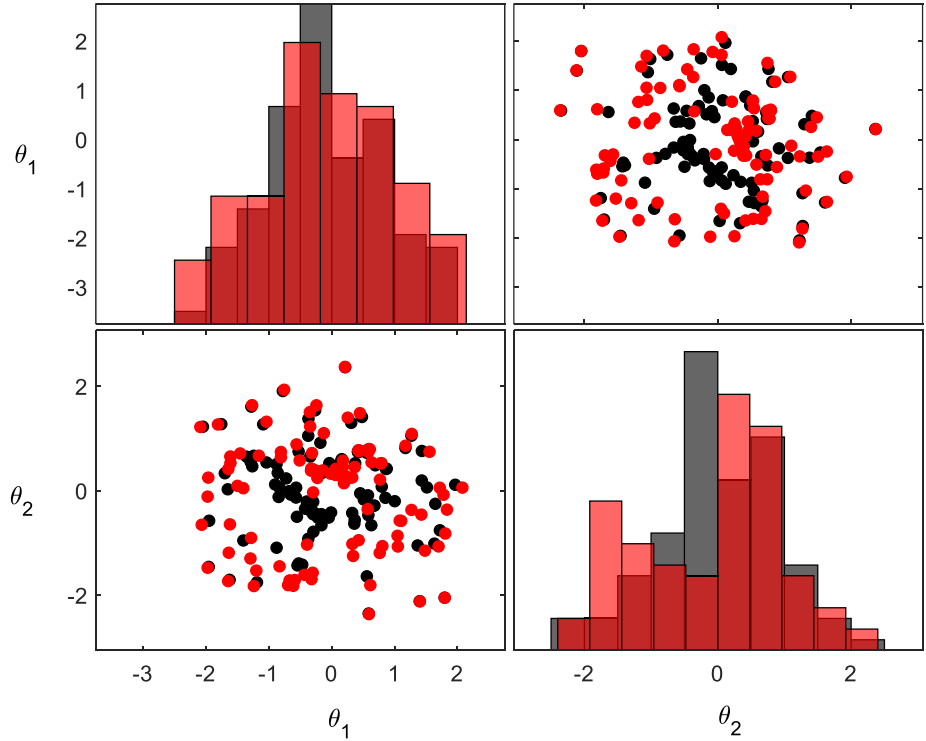

Figure 17. Scatterplots and histograms show the prior, black – optimal prior case; red – worst-case prior.

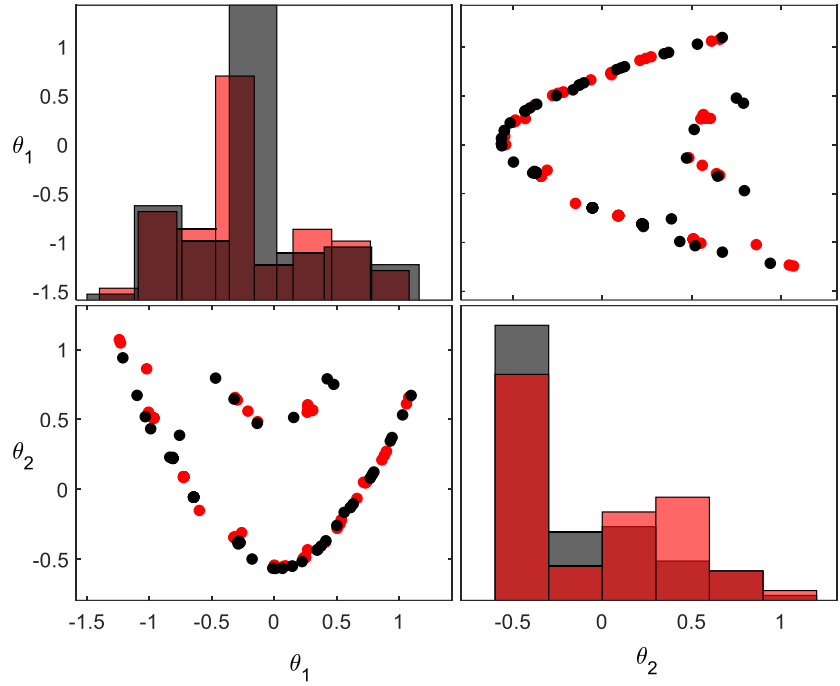

Figure 18. Scatterplots and histograms show the approximation to the posterior, black – optimal prior case; red – worst-case prior.

### 6.2. Double beam structure example

The model used in this second example is based on the coupled beam structure illustrated on the first version of the preprint (Igea & Cicirello, 2022).

The structure is shown on Fig.19, two connecting fixtures composed of three springs each: one translational, one shear, and one rotational, link two beams. This example shows practical interest, as it can be used to depict structural conditions where the attaching ensembles between elements show uncertainty. The causes of such uncertainty can be derived from boundary conditions and manufacturing variability. More specifically, the four uncertain parameters chosen are the spring stiffnesses and the young's modulus of both beams: the rotational springs $k_2 = 500\theta_1$ [Nm/rad], the shear springs $k_3 = 10^7 \theta_2$ [N/m], the translational springs $k_1 = 10^{10}\theta_3$ [N/m] and the Young's modulus of both beams $E_1 = E_2 = 210*10^9 \theta_3$ [Pa]. For those four uncertain parameters, the initial prior distribution is a multivariate gaussian prior chosen as $\mathcal{N}(I, 0.03I)$.

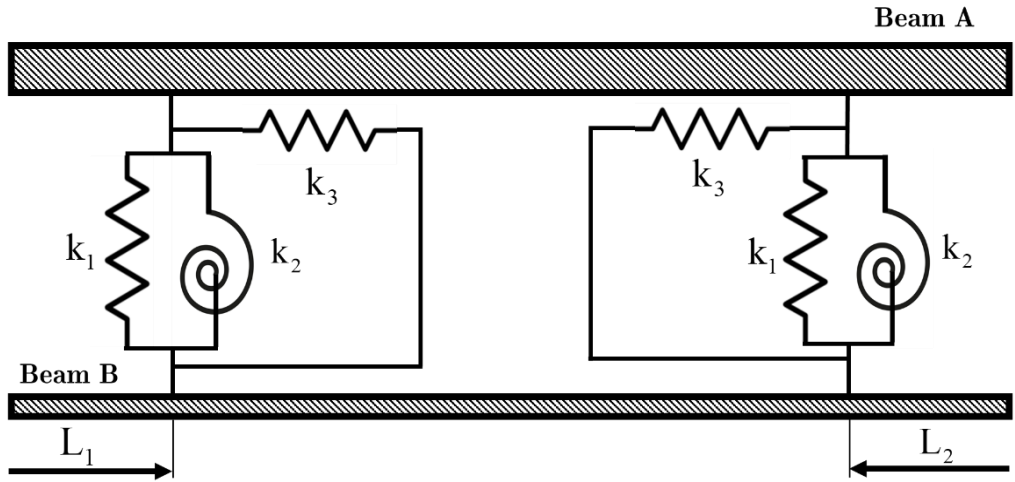

Figure 19. Theoretical model of a coupled beam structure.

Dimensions and mechanical characteristics of the double beam model may be found on Table 1.

**Table 1**

Coupled beam dimensions, distances from edges to connections, and mechanical characteristics.

|  | Thickness | Width | Length | $L_1$ | $L_2$ | Young's modulus | Density |
|---|---|---|---|---|---|---|---|
|  |  |  | [mm] |  |  | [GPa] | [Kg/m³] |
| Beam A | 6 | 25 | 600 | 20 | 20 | 210 | 7800 |
| Beam B | 3 |  |  |  |  |  |  |

| Springs | $k_1$ | $k_3$ | $k_2$ |
|---|---|---|---|
|  | [MN/m] |  | [Nm/rad] |
|  | 100 | 10 | 500 |

Using the data on Table 1, the first eight natural frequencies of the model were assessed and introduced on Table 2.

**Table 2**

Coupled beam structure natural frequencies [Hz].

| $f_1$ | $f_2$ | $f_3$ | $f_4$ | $f_5$ | $f_6$ | $f_7$ | $f_8$ |
|---|---|---|---|---|---|---|---|
| 16.0 | 50.2 | 92.8 | 134.6 | 245.3 | 260.7 | 428.0 | 478.6 |

The numerical frequencies obtained on Table 2 were produced using a Finite Element (FE) code. The code assumes a 2-dimensional Euler-Bernoulli beam model. Uniform discretization with two hundred FEs for each beam was used. Each FE has two nodes and each node has two degrees of freedom.

The likelihood function is assumed to be a multivariate gaussian distribution, the mean is given by the deterministic value of the eight natural frequencies in Table 2, and the covariance is assumed to be diagonal covariance matrix that has standard deviations of 2% of their deterministic values ($\sigma_i = 0.02 f_i$).

In this example, for the definition of the ambiguity set, the statistical distance used is also the 2-Wassertein distance, where the radius is $\varepsilon = 0.04$, and a nominal prior equal to the initial prior is selected.

The values described above are used as inputs of the algorithm, and the interacting Wasserstein gradient flows are used to find the resulting distributions for two different cases: a) the optimal prior and its resulting approximation to the posterior, and b) the worst-case prior and its resulting approximation to the posterior. In this example, the gaussian process method described in section 5.2.2 is used to calculate an approximation to the gradient of the log likelihood at the particle positions evaluated.

The values of step size used in the interacting particle flow WGF algorithm for the optimal prior case are $\alpha = 5*10^{-5}$ and $\tau = 2.5*10^{-3}$. The values used for the worst-case prior case are $\alpha = 5*10^{-5}$ and $\tau = 5*10^{-5}$.

Figures 20 and 21 respectively illustrate the positions of the particles from the optimal prior and from the approximation to the posterior for each iteration. In Figure 20, after iteration $i = N_a$, it can be seen that for $\theta_1$, $\theta_2$ and $\theta_4$ the prior particle positions start concentrating at values close to one. It can also be seen that $\theta_4$ has the most rapid change out of all the latent variables, this is probably due to being the latent variable which most affects the model output. However, the opposite effect can be observed for $\theta_3$, this is most likely due to the low sensitivity of the model output to changes of the latent variable $\theta_3$. Figure 21 shows how the particles of the approximation to the posterior also concentrate to values closer to one as the number of iterations progresses for all the latent variables except for $\theta_3$.

Figure 22 show scatter plots and histogram produced by the `plotmatrix` function of (*MATLAB*, 2020), of the initial particles from the prior/approximation to the posterior (red), the final particles from the optimal prior (black) and the final particles from approximation to the posterior (blue). It can be seen that the particles positions from the optimal prior and the approximation to the posterior are quite similar for all latent variables except for $\theta_3$. From the histogram, it can be also seen that for the latent variable $\theta_3$, the optimal prior has a bigger support than the initial prior.

The positions of the particles from the worst-case prior, and from the approximation to the posterior for each iteration are shown on Figures 23 and 24 respectively. Figure 23 shows that after iteration $i = N_a$ for $\theta_1$, $\theta_2$ and $\theta_4$ the prior particle positions part from values close to one. This is the opposite of what occurs for the optimal prior case. In a manner similar to what happens for the optimal case, the latent variable $\theta_4$ experiments the fastest change of all the uncertain parameters. This is most likely due to the higher sensitivity of the model output to the changes of this latent variable. Figure 24 illustrates how as the number of iterations progresses the particles of the approximation to the posterior depart from values close to one. However, in this case, the change in the positions of the particles of the approximation to the posterior is not as significant as in the case for the optimal prior.

Histograms and scatter plots produced by the `plotmatrix` function of (*MATLAB*, 2020), can be found on Figure 25. The graphs illustrate the positions of the particles. In blue, the final particles from approximation to the posterior. In black, the final particles from the optimal prior. In red, the initial particles from the prior/approximation to the posterior. From the histograms, it can be deduced that in the worst-case prior, the supports of the graphs of all latent variables are bigger compared to the ones of the initial prior. The scatterplots show that worst-case prior particles have moved in such a manner that most of its particles lie in regions of lower posterior density.

Figure 26 directly compares the optimal prior and worst-case prior using the `plotmatrix` function from (*MATLAB*, 2020) by plotting scatter plots and the histograms of the latent variables. In general, for most of the latent variables it can be seen that the support of the worst-case prior is bigger than the optimal prior.

Figure 27 also has scatter plots and the histograms of the latent variables of the resulting approximation to the posterior when the optimal prior and worst-case prior has been calculated. When comparing the resulting approximations to the posteriors, it can be seen that for the case with the optimal prior the resulting approximation to the posterior is more concentrated compared to the approximation to the posterior that results from the worst-case prior. In this example, it is seen that the posterior is slightly sensitive to the considered uncertain prior.

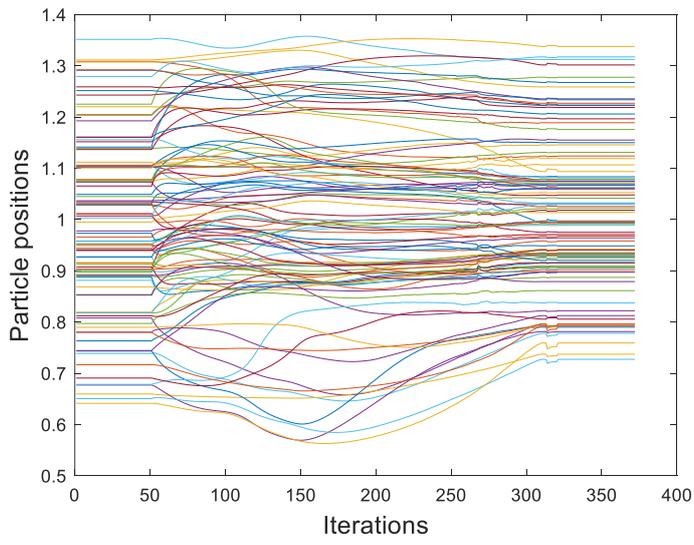
(a)

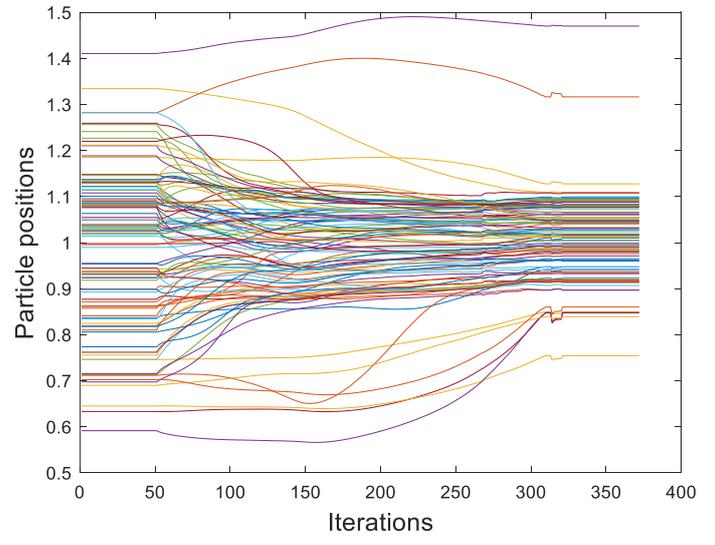
(b)

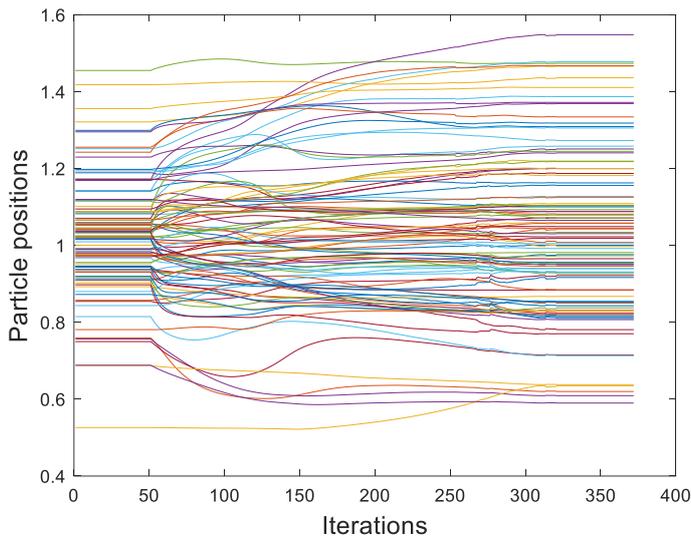
(c)

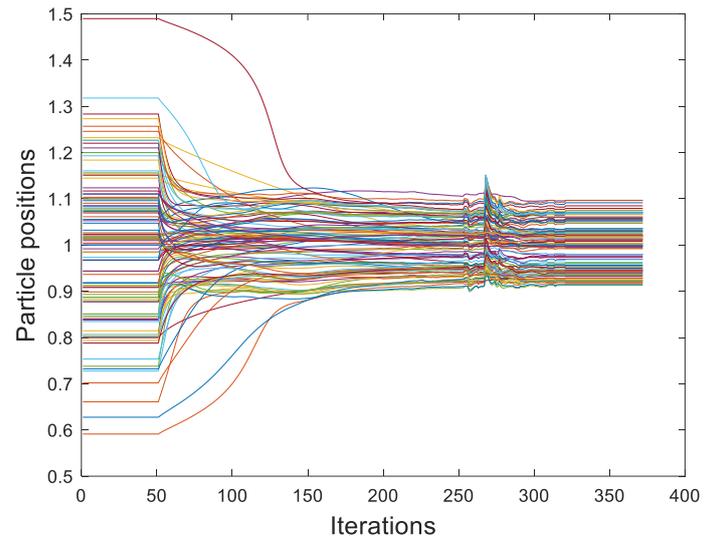
(d)

Figure 20. Optimal prior particle positions at different iterations for different latent parameters: a) $\theta_1$; b) $\theta_2$; c) $\theta_3$; d) $\theta_4$.

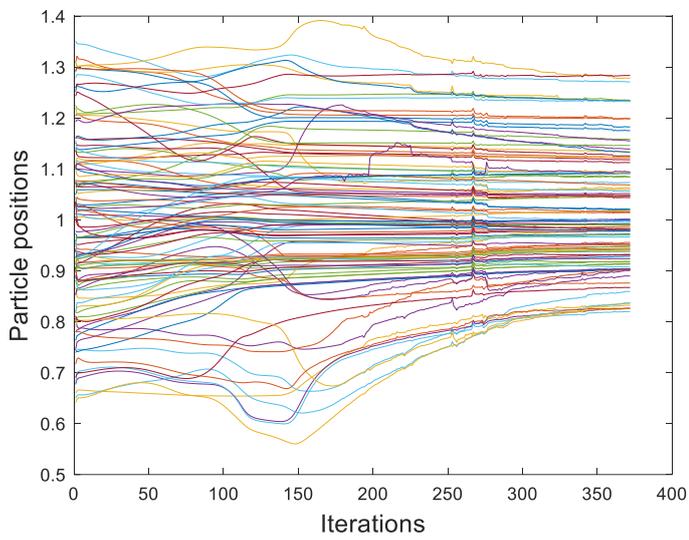

(a)

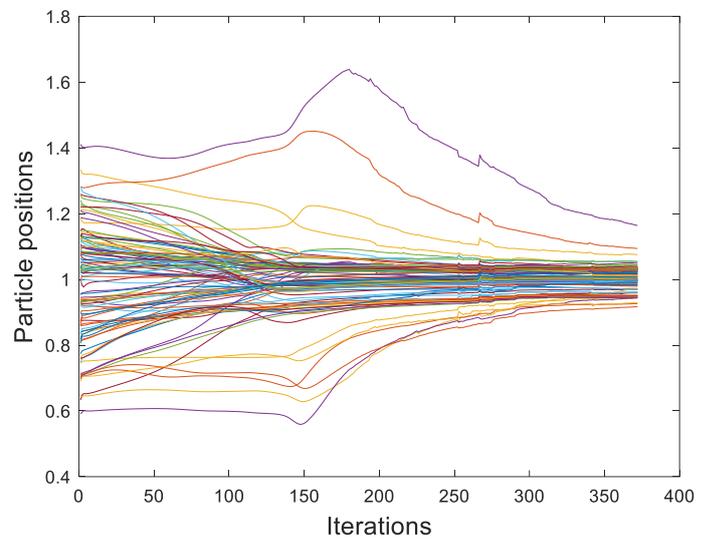

(b)

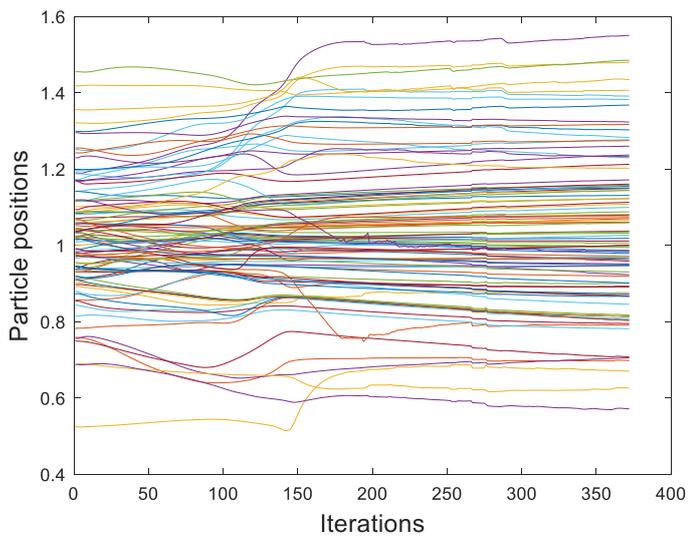

(c)

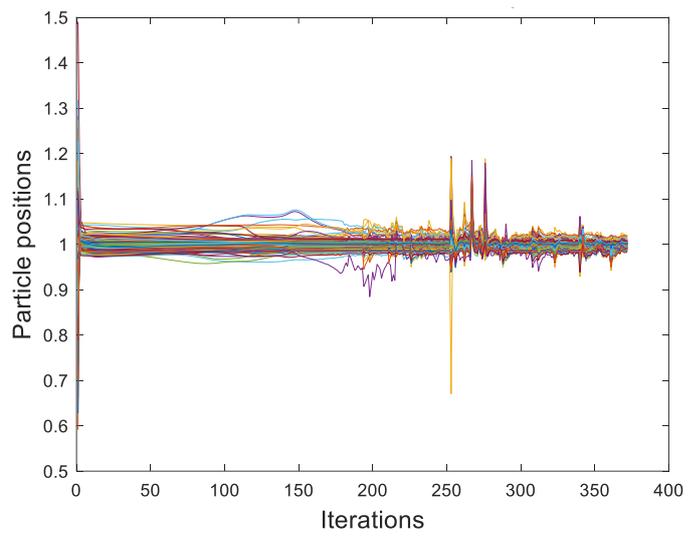

(d)

Figure 21. Approximation to posterior particle positions at different iterations for different latent parameters: : a) $\theta_1$; b) $\theta_2$; c) $\theta_3$; d) $\theta_4$.

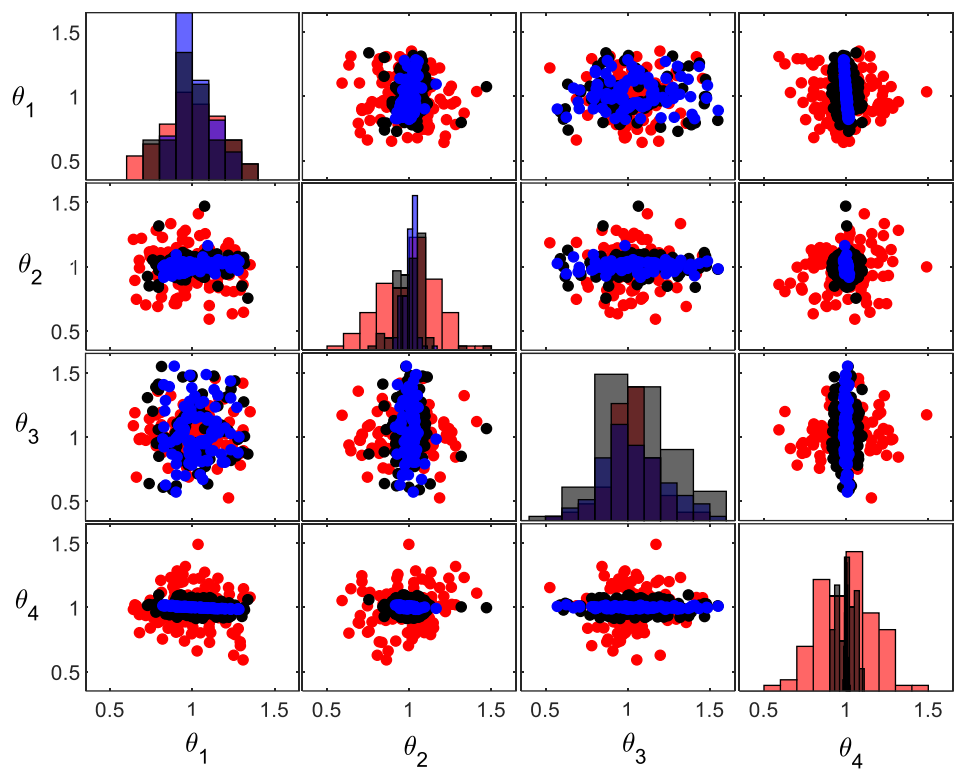

Figure 22. Scatterplots and histograms show: red – initial prior; blue – final approximation to the posterior; black – optimal prior.

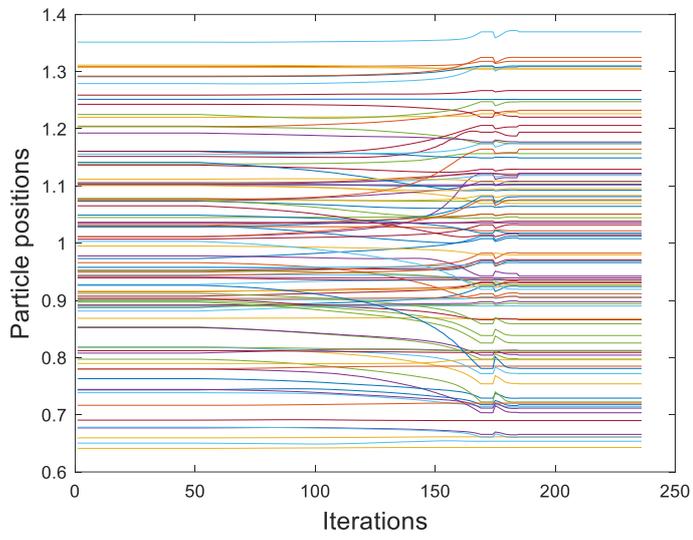
(a)

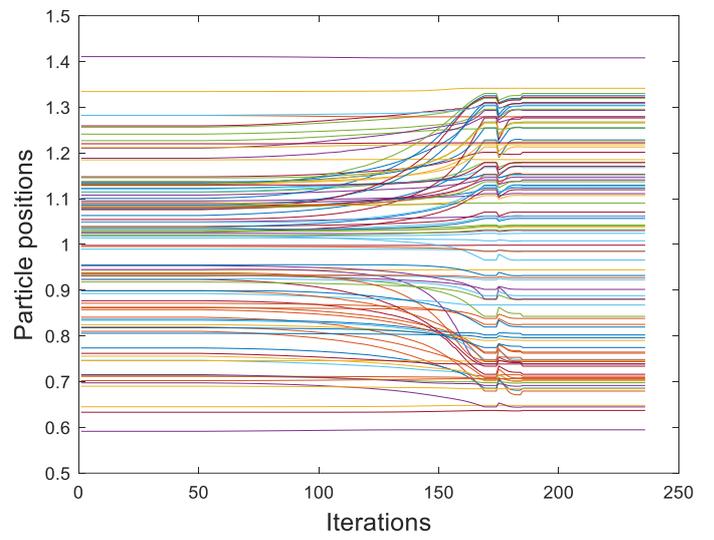
(b)

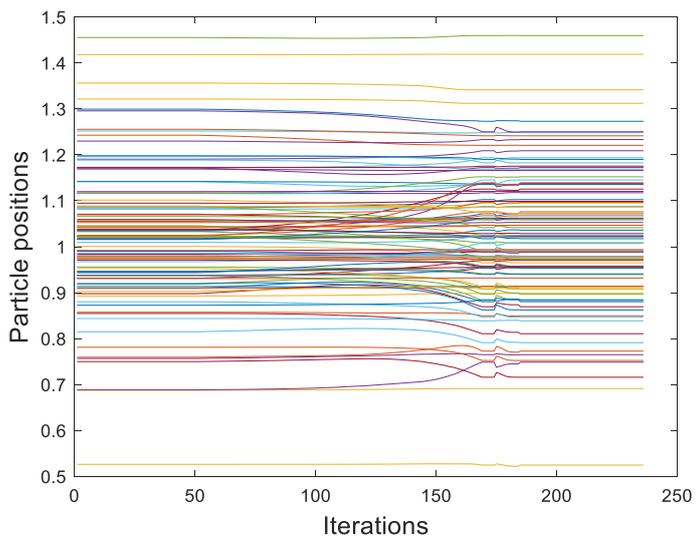
(c)

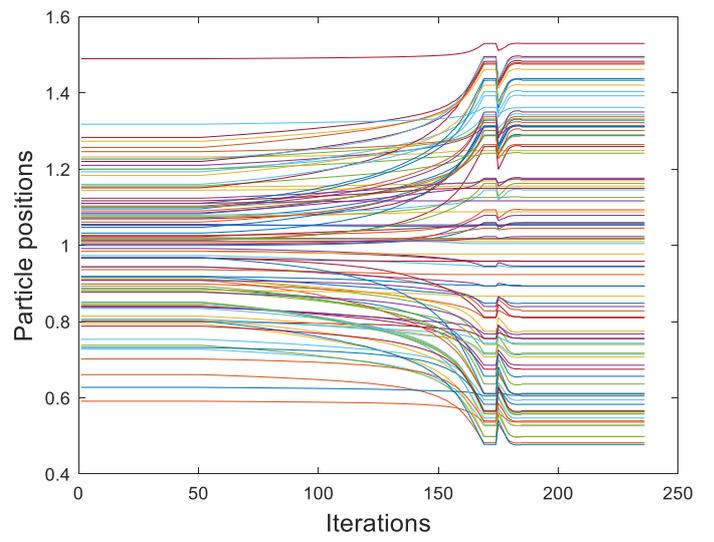
(d)

Figure 23. Worst-case prior particle positions at different iterations for different latent parameters: : a) $\theta_1$; b) $\theta_2$; c) $\theta_3$; d) $\theta_4$.

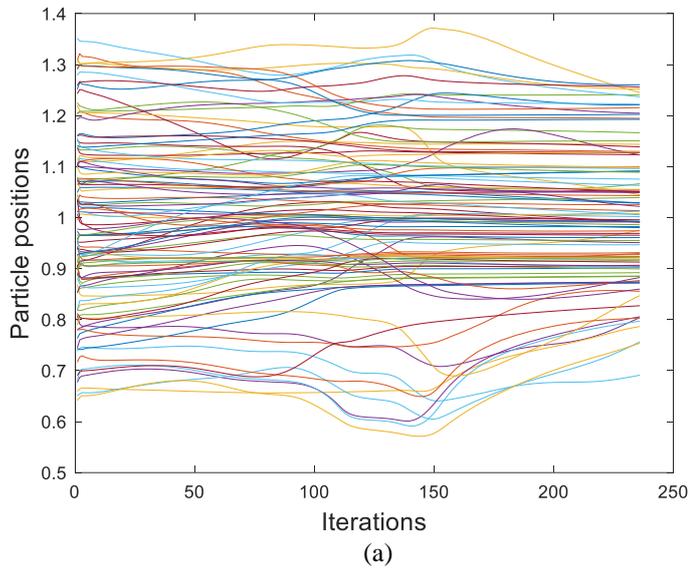
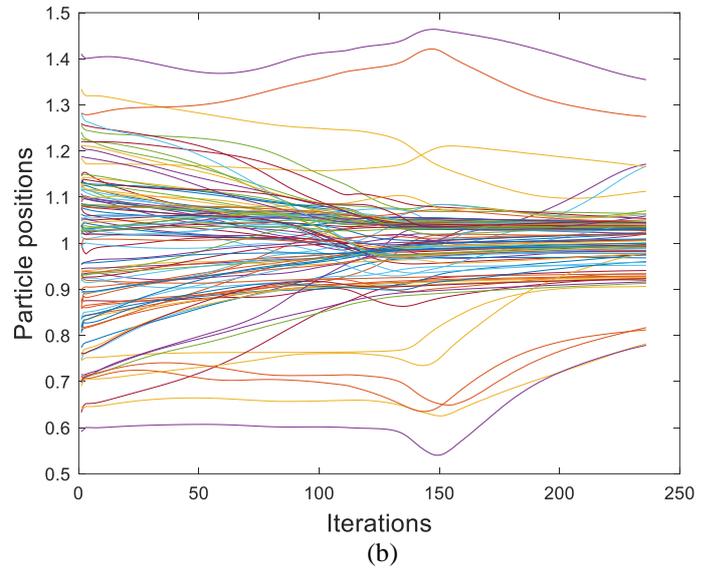
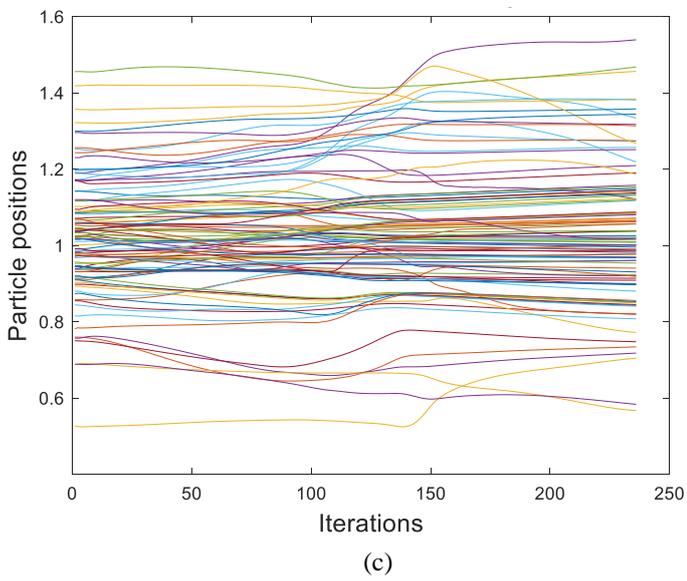
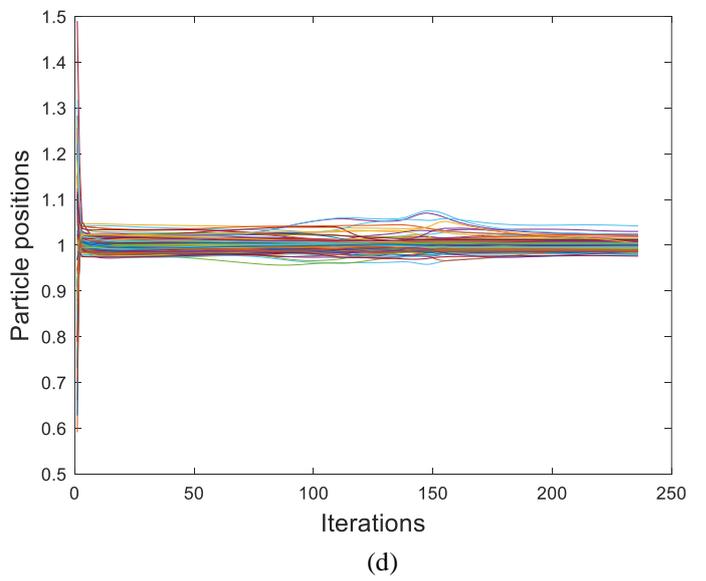

Figure 24. Approximation to posterior particle positions at different iterations for different latent parameters: a) $\theta_1$; b) $\theta_2$; c) $\theta_3$; d) $\theta_4$.

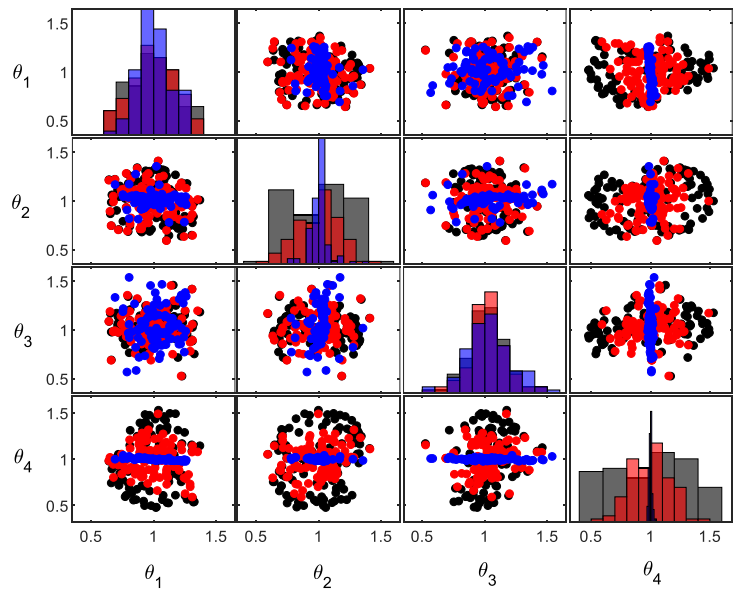

Figure 25. Scatterplots and histograms show: red – initial prior; blue – final approximation to the posterior; black – worst-case prior.

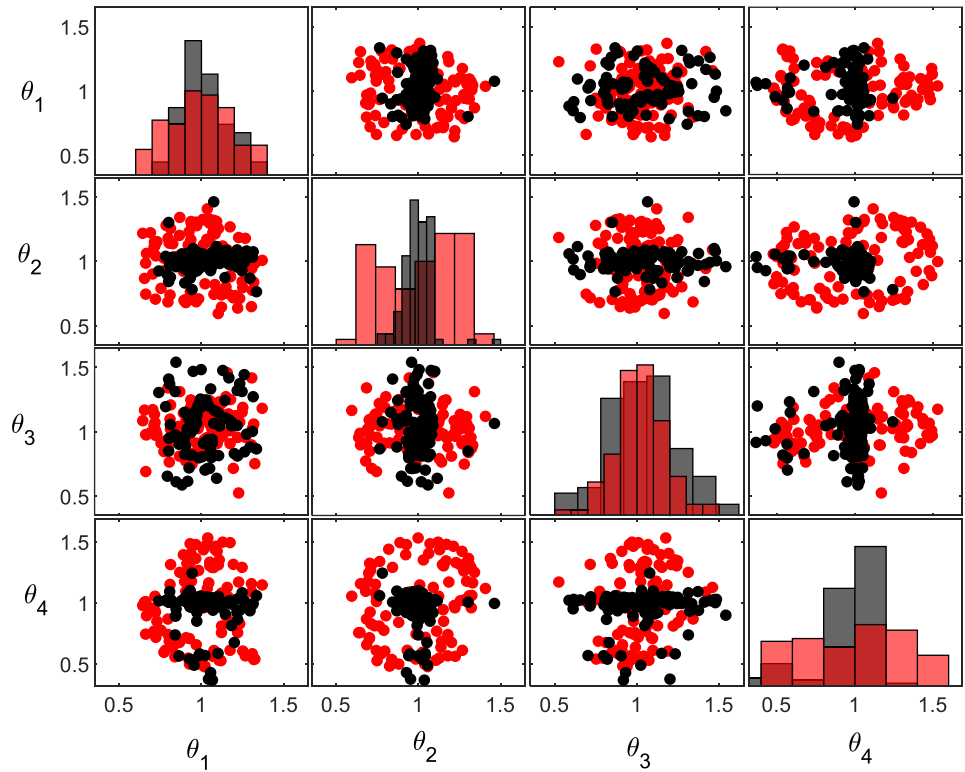

Figure 26. Scatterplots and histograms show the prior, black – optimal prior case; red – worst-case prior.

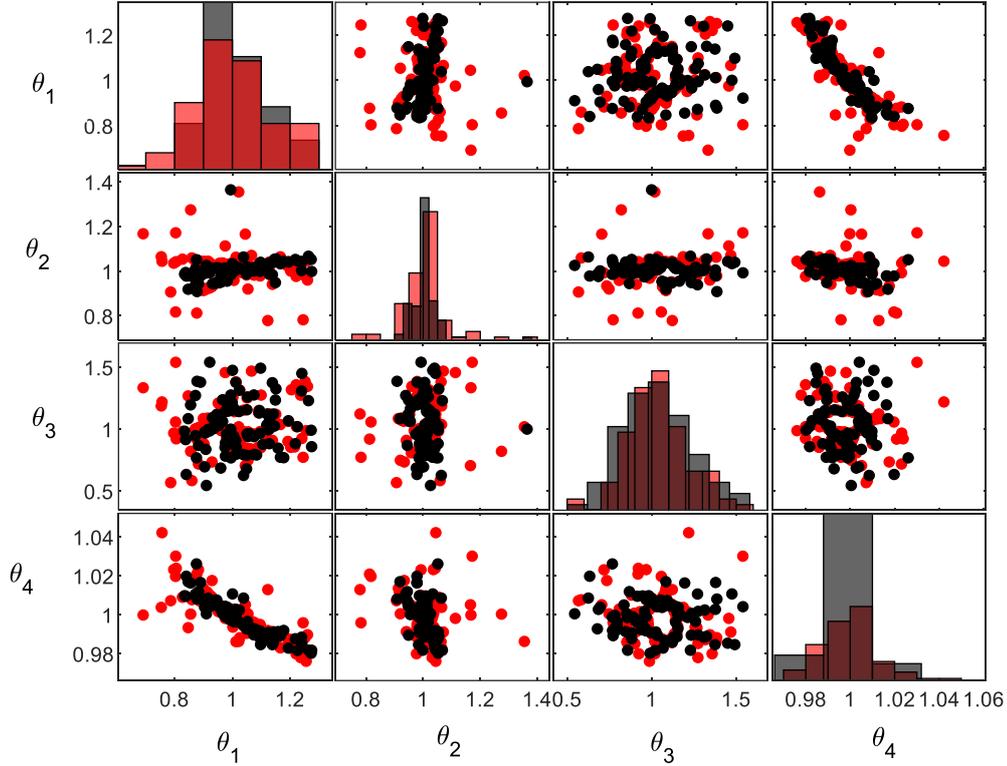

Figure 27. Scatterplots and histograms show the approximation to the posterior, black – optimal prior case; red – worst-case prior.

## 7. Conclusion

In this paper, a Robust Bayesian Inference approach, based on Wasserstein gradient flows for the robust estimation of the latent parameters' posterior of a physics-based model, given observed data, has been proposed. The proposed approach yields an estimation of the posterior distribution of the identified parameters by finding the optimal and worst-case prior distribution. This estimation is produced by an algorithm that combines an interacting Wasserstein gradient flow formulation with an ambiguity set. The ambiguity set is defined by a nominal distribution, a statistical distance and a radius. In this paper, the 2-Wasserstein distance is used as the statistical distance. Due to the properties of the 2-Wasserstein distance, this means that the distributions that lie inside the prescribed radius do not need to have the same support. The ambiguity set may be used to explore the sensitivity of the posterior prediction of the system to uncertainty in the prior. This investigation may be of particular interest for cases where the opinions of different experts are conflicting.

The interacting Wasserstein gradient flow formulation is derived from first principles, obtaining particle discretisation equations for the calculation of the optimal and worst-case prior. The derivation of the interacting WGFs, allows the development of the proposed method, which may reduce the computational cost incurred if all the possible prior distributions that lie inside the ambiguity set were to be tested directly. A kernel density estimator is used to obtain estimates of the gradient of the logarithm of the prior, and of the gradient of the logarithm of the approximations to the posterior with respect to the particle positions.

The paper illustrates how the gradient of the logarithm of the likelihood may be estimated either using an ensemble method, or a gaussian process regression method. Two numerical examples have been used to show both the optimal and worst-case prior and their resulting approximation to the posterior. In these examples, it is shown that for the optimal prior case the particles' positions tend to be near the particles of the approximation to the posterior, this means the optimal prior assigns a higher prior density close to regions of high posterior density.

For the worst-case prior, the opposite behaviour may be seen, the particles tend to move to positions far from the particles of the approximation to the posterior density. As a consequence, the worst-case prior has a bigger support than the initial prior.

The proposed approach may be of application to areas outside the scope of structural engineering, with particular interest to the commonly occurring engineering cases where some of the latent parameters may exhibit higher uncertainty than others.

Future work may focus on the convergence properties of the proposed approach and the dynamic selection of step sizes, as this would allow the proposed approach to become more computationally efficient. The method would also benefit from the development of a sample-efficient strategy, in which the reuse of samples from previous iterations may be integrated into the proposed methodology, reducing the number of simulations further. Another potential direction of interest is the development of a principled approach for the selection of the nominal prior and its radius, as at this stage is assumed to be known. These topics are currently under investigation.

**Funding Statement**

Felipe Igea and Alice Cicirello thank the EPSRC and Schlumberger for an Industrial Case postgraduate scholarship (Grant ref. EP/T517653/1). Alice Cicirello gratefully acknowledges the financial support provided by the Alexander von Humboldt Foundation Research Fellowship for Experienced Researchers supporting part of this research.

**Competing Interests**

None.

**Data Availability Statement**

The data that support the findings of this study will be made available upon request.

**References**

Ambrosio, Luigi., Gigli, Nicola., & Savaré, Giuseppe. (2005). *Gradient flows : in metric spaces and in the space of probability measures*. Birkhäuser.

Bayraksan, G., & Love, D. K. (2015). Data-Driven Stochastic Programming Using Phi-Divergences. *INFORMS Tutorials in Operations Research*, 1–19. https://doi.org/10.1287/EDUC.2015.0134

Blei, D. M., Kucukelbir, A., & McAuliffe, J. D. (2017). Variational Inference: A Review for Statisticians. In *Journal of the American Statistical Association* (Vol. 112, Issue 518, pp. 859–877). American Statistical Association. https://doi.org/10.1080/01621459.2017.1285773

Chen, C., Zhang, R., Wang, W., Li, B., & Chen, L. (2018). A Unified Particle-Optimization Framework for Scalable Bayesian Sampling. *34th Conference on Uncertainty in Artificial Intelligence 2018, UAI 2018*, *2*, 746–755. https://arxiv.org/abs/1805.11659v2

Chen, Y., Huang, D. Z., Huang, J., Reich, S., & Stuart, A. M. (2023). *Gradient Flows for Sampling: Mean-Field Models, Gaussian Approximations and Affine Invariance*. https://arxiv.org/abs/2302.11024v3

Chérief-Abdellatif, B.-E., & Alquier, P. (2019). *MMD-Bayes: Robust Bayesian Estimation via Maximum Mean Discrepancy*. https://arxiv.org/abs/1909.13339v2

Chizat, L., & Bach, F. (2018). On the Global Convergence of Gradient Descent for Over-parameterized Models using Optimal Transport. *Advances in Neural Information Processing Systems*, *2018-December*, 3036–3046. https://arxiv.org/abs/1805.09545v2

Detommaso, G., Cui, T., Spantini, A., Marzouk, Y., & Scheichl, R. (2018). A Stein variational Newton method. *Advances in Neural Information Processing Systems*, *2018-December*, 9169–9179. https://arxiv.org/abs/1806.03085v2

Dewaskar, M., Tosh, C., Knoblauch, J., & Dunson, D. B. (2023). *Robustifying likelihoods by optimistically re-weighting data*. https://arxiv.org/abs/2303.10525v1

Diakonikolas, I., Hopkins, S. B., Kane, D., & Karmalkar, S. (2020). *Robustly Learning any Clusterable Mixture of Gaussians*. https://arxiv.org/abs/2005.06417v1

Ding, S., Dong, H., Fang, C., Lin, Z., & Zhang, T. (2023). *Provable Particle-based Primal-Dual Algorithm for Mixed Nash Equilibrium*. https://arxiv.org/abs/2303.00970v1

Dunbar, O. R. A., Duncan, A. B., Stuart, A. M., & Wolfram, M. T. (2022). Ensemble Inference Methods for Models With Noisy and Expensive Likelihoods. *Https://Doi.Org/10.1137/21M1410853*, *21*(2), 1539–1572. https://doi.org/10.1137/21M1410853

Frazier, D. T. (2020). *Robust and Efficient Approximate Bayesian Computation: A Minimum Distance Approach*. https://arxiv.org/abs/2006.14126v1

Gao, Y., & Liu, J.-G. (2020). A note on parametric Bayesian inference via gradient flows. *Annals of Mathematical Sciences and Applications*, *5*(2), 261–282.

Ghosh, A., & Basu, A. (2016). Robust Bayes estimation using the density power divergence. *Annals of the Institute of Statistical Mathematics*, *68*(2), 413–437. https://doi.org/10.1007/S10463-014-0499-0/FIGURES/10

Go, J., & Isaac, T. (2022). Robust Expected Information Gain for Optimal Bayesian Experimental Design Using Ambiguity Sets. *Proceedings of the 38th Conference on Uncertainty in Artificial Intelligence, UAI 2022*, 728–737. https://arxiv.org/abs/2205.09914v1

Gonçalves, F. B., Prates, M. O., & Lachos, V. H. (2015). Robust Bayesian model selection for heavy-tailed linear regression using finite mixtures. *Brazilian Journal of Probability and Statistics*, *34*(1), 51–70. https://doi.org/10.1214/18-BJPS417

Hermans, J., Delaunoy, A., Rozet, F., Wehenkel, A., Begy, V., & Louppe, G. (2021). *A Trust Crisis In Simulation-Based Inference? Your Posterior Approximations Can Be Unfaithful*. https://arxiv.org/abs/2110.06581v3

Hooker, G., & Vidyashankar, A. N. (2011). Bayesian Model Robustness via Disparities. *Test*, *23*(3), 556–584. https://doi.org/10.1007/s11749-014-0360-z

Igea, F., & Cicirello, A. (2022). Cyclical Variational Bayes Monte Carlo for Efficient Multi-Modal Posterior Distributions Evaluation. In *Mechanical Systems and Signal Processing* (Vol. 186). Academic Press. http://arxiv.org/abs/2202.11645v1

Kuhn, D., Esfahani, P. M., Nguyen, V. A., & Shafieezadeh-Abadeh, S. (2019). Wasserstein Distributionally Robust Optimization: Theory and Applications in Machine Learning. *Operations Research & Management Science in the Age of Analytics*, 130–166. https://doi.org/10.1287/educ.2019.0198

Lin, T., Jin, C., & Jordan, M. I. (2019). On Gradient Descent Ascent for Nonconvex-Concave Minimax Problems. *37th International Conference on Machine Learning, ICML 2020*, *PartF168147-8*, 6039–6049. https://arxiv.org/abs/1906.00331v8

Liu, C., Cheng, P., Zhang, R., Zhuo, J., Zhu, J., & Carin, L. (2018). *Accelerated First-order Methods on the Wasserstein Space for Bayesian Inference*. https://www.researchgate.net/publication/326222852

Liu, Q., & Wang, D. (2016). Stein Variational Gradient Descent: A General Purpose Bayesian Inference Algorithm. *Advances in Neural Information Processing Systems*, 2378–2386. https://arxiv.org/abs/1608.04471v3

Lu, J., Lu, Y., & Nolen, J. (2019). Scaling Limit of the Stein Variational Gradient Descent: The Mean Field Regime. *Https://Doi.Org/10.1137/18M1187611*, *51*(2), 648–671. https://doi.org/10.1137/18M1187611

Lu, Y. (2022). *Two-Scale Gradient Descent Ascent Dynamics Finds Mixed Nash Equilibria of Continuous Games: A Mean-Field Perspective*. https://arxiv.org/abs/2212.08791v2

Lyddon, S., Walker, S., & Holmes, C. (2018). Nonparametric learning from Bayesian models with randomized objective functions. *Advances in Neural Information Processing Systems*, *2018-December*, 2071–2081. https://arxiv.org/abs/1806.11544v2

*MATLAB* (a). (2020).

Matsubara, T., Knoblauch, J., Briol, F. X., & Oates, C. J. (2021). Robust Generalised Bayesian Inference for Intractable Likelihoods. *Journal of the Royal Statistical Society. Series B: Statistical Methodology*, *84*(3), 997–1022. https://doi.org/10.1111/rssb.12500

Mei, S., Montanari, A., & Nguyen, P. M. (2018). A mean field view of the landscape of two-layer neural networks. *Proceedings of the National Academy of Sciences of the United States of America*, *115*(33), E7665–E7671. https://doi.org/10.1073/PNAS.1806579115/SUPPL_FILE/PNAS.1806579115.SAPP.PDF

Ramgraber, M., Weatherl, R., Blumensaat, F., & Schirmer, M. (2021). Non-Gaussian Parameter Inference for Hydrogeological Models Using Stein Variational Gradient Descent. *Water Resources Research*, *57*(4), e2020WR029339. https://doi.org/10.1029/2020WR029339

Rasmussen, C. E. (2003). Gaussian Processes to Speed up Hybrid Monte Carlo for Expensive Bayesian Integrals. *BAYESIAN STATISTICS*, *7*, 651–659.

Santambrogio, F. (2015). *Optimal Transport for Applied Mathematicians* (Vol. 87). Springer International Publishing. https://doi.org/10.1007/978-3-319-20828-2

Santambrogio, F. (2016). { Euclidean, Metric, and Wasserstein } Gradient Flows: an overview. *Bulletin of Mathematical Sciences*, *7*(1), 87–154. https://doi.org/10.1007/s13373-017-0101-1

van Parys, B. P. G., Esfahani, P. M., & Kuhn, D. (2017). From Data to Decisions: Distributionally Robust Optimization is Optimal. *Management Science*, *67*(6), 3387–3402. https://doi.org/10.1287/mnsc.2020.3678

Wang, Y., Chen, P., & Li, W. (2022). Projected Wasserstein Gradient Descent for High-Dimensional Bayesian Inference. *Https://Doi.Org/10.1137/21M1454018*, *10*(4), 1513–1532. https://doi.org/10.1137/21M1454018

Yamada, M., Suzuki, T., Kanamori, T., Hachiya, H., & Sugiyama, M. (2011). Relative Density-Ratio Estimation for Robust Distribution Comparison. *Neural Computation*, *25*(5), 1324–1370. https://doi.org/10.1162/NECO_a_00442